\documentclass[final,5p,twocolumn]{elsarticle}

\usepackage{lineno}
\usepackage{hyperref,color,amsmath}
\usepackage{subfigure, float}

\modulolinenumbers[5]

\renewcommand{\vec}[1]{\ensuremath\boldsymbol{#1}}

\journal{Computer Physics Communications}

\begin{document}

\begin{frontmatter}

\title{A General, Mass-Preserving Navier-Stokes Projection Method}

\author{David Salac}
\address{Mechanical and Aerospace Engineering, University at Buffalo SUNY, 318 Jarvis Hall, Buffalo, NY 14260-4400, USA}

\begin{abstract}

The conservation of mass is common issue with multiphase fluid simulations. In this work a novel projection method is presented
which conserves mass both locally and globally. The fluid pressure is augmented with a time-varying component which accounts
for any global mass change. The resulting system of equations is solved using an efficient Schur-complement method. 
Using the proposed method four numerical examples are performed: the evolution of a static bubble, the rise of a bubble,
the breakup of a thin fluid thread, and the extension of a droplet in shear flow. The method is capable of conserving 
the mass even in situations with morphological changes such as droplet breakup.

\end{abstract}


\end{frontmatter}


\section{Introduction}
A wide variety of techniques have been proposed to model multiphase fluid systems with incompressible, immiscible fluids.
These include explicit front-tracking techniques \cite{UNVERDI1992, Tryggvason2001}, the volume-of-fluid method \cite{Weymouth2010, HIRT1981},
the phase-field method \cite{Jacqmin1999, Anderson1998, Aland2010}, 
and the level set method \cite{Shin2005, Chang1996}. 
A constant challenge in each of these techniques is the conservation of mass during the course of the simulation.
Each of these methods handle this challenge differently. For example, the volume-of-fluid methods has excellent mass conservation
properties \cite{Baraldi2014} at the expense of requiring complex 
heuristic interface reconstructions techniques to calculate geometric quantities such as curvature \cite{Cummins2005,Pilliod2004,Raessi2007}. 

Unfortunately, there are many physical systems where high accuracy of geometric quantities are required. One example are models of liposome vesicles 
where interfacial forces depend on high order derivatives of the interface's curvature \cite{Salac2011,Kolahdouz2015b,Yazdani2012}.
Unlike volume-of-fluid techniques, front tracking, phase-field, and level-set methods are able to provide higher geometric accuracy.
This accuracy is obtained at the expense of natural volume conservation and thus special care must be taken to ensure
that mass does not change over the course of a simulation. 

There have been numerous attempts to improve the mass conservation of such methods. For example, level set methods have 
been adjusted by Lagrangian particles
which are used to correct the level set function \cite{Enright2005,Enright2002} or level sets have been combined with
volume of fluid methods \cite{Sussman2000}. Other simply shift the interface function to match the volume constraint \cite{Yap2006}.

\begin{figure}[!ht]
	\begin{center}
		\includegraphics[width=1.5in]{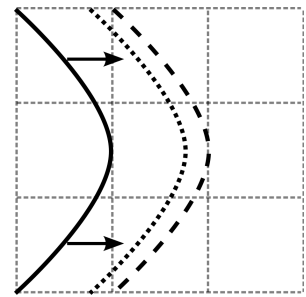}		
		\caption{Schematic of velocity-interface decoupling. The interface begins on the left (solid line). The underlying flow field dictates that the interface
		move to the right and should end as the dash-line after a single time step. Due to an externally applied correction the interface instead ends the time step
		at the dotted line.}
		\label{fig:velocitydecouple}
	\end{center}
\end{figure}

The major issue with these types of corrections is that the interface becomes decoupled from the underlying flow field. As an example consider
an interface on a uniform Cartesian grid, Fig. \ref{fig:velocitydecouple}. The interface begins on the left side and the underlying fluid flow-field
dictates that it should move to the right one grid spacing. After this time step it is determined that errors in the simulation 
resulted in a mass gain. A simple and often-used correction is to simply move the interface to obtain mass conservation. In this case the effect is 
the interface will not move the full amount that the underlying flow field dictates. The movement of the interface and the underlying flow field
have become decoupled. Any externally applied correction for front-tracking, phase-field or level set methods will demonstrate similar behavior.

In this manuscript a different approach is taken. Instead of adjusting the interface to achieve mass conservation a novel
Navier-Stokes projection method is developed which ensures mass conservation. Unlike the previous method the interface is simply advected
due to the underlying flow-field, it is the flow-field itself which explicitly takes into account any possible mass-loss. As will
become apparent later in the manuscript it is useful to think of this method as modifying the pressure so that it can handle not only
local incompressibility but also global mass conservation. Note that while this manuscript will focus on a particular Navier-Stokes
numerical implementation, the concept presented here extends to any type of multiphase fluid simulation that uses a projection method.

The remainder of the manuscript is as follows. In Section \ref{sec:single-fluid} the single-fluid formulation of multiphase fluid flow
is briefly described. The novel mass-preserving Navier-Stokes projection method is described in Section \ref{sec:projection-method}.
The numerical implementation is given in Section \ref{sec:numerics}, which is followed by numerical experiments in Section \ref{sec:experiments}.
A short conclusion is presented in Section \ref{sec:conclusion}.

\section{Single-Fluid Navier-Stokes Equations}
\label{sec:single-fluid}
\begin{figure}[!ht]
	\begin{center}
		\includegraphics[width=2in]{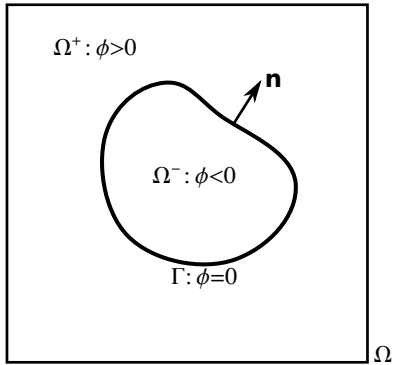}		
		\caption{The level-set description of a multiphase fluid system. The outward facing normal is shown for clarity.}
		\label{fig:levelset}
	\end{center}
\end{figure}

In this sections a brief introduction to the single-fluid formulation of multiphase fluid flow is presented. In a multiphase fluid system
the interface between two immiscible fluids evolves over time. In this work a level-set description of the interface is used. Let the 
evolving interface be given as the set of points where the level-set function is zero, $\Gamma(t)=\{\vec{x}:\phi(\vec{x},t)=0\}$.
The evolution of the level set function $\phi(\vec{x},t)$ over time will implicitly determine the location of the interface.
Following convention the interior fluid, $\Omega^-$, is given by $\phi<0$ while the outer fluid, $\Omega^+$, is given by $\phi>0$, Fig. \ref{fig:levelset}.
The entire domain is given as $\Omega=\Omega^+\cup\Omega^-$. Using the level-set description it is possible to obtain geometric
information of the interface easily. For example, the outward facing normal is simply $\vec{n}=\nabla\phi/\|\nabla\phi\|$ while the total curvature can be calculated
as $\kappa=\nabla\cdot\vec{n}$.

In each domain the Navier-Stokes equations hold,
\begin{align}
	\label{eq:NS} \rho^\pm\dfrac{D\vec{u}^\pm}{Dt}&=-\nabla p^\pm+\nabla\cdot\left(\mu^\pm\left(\nabla\vec{u}^\pm+\nabla^T\vec{u}^\pm\right) \right)+\vec{g}^\pm, \\
	\label{eq:DIV} \nabla\cdot\vec{u}^\pm&=0,
\end{align}
where $\rho$ is the fluid density, $\mu$ is the fluid viscosity and $\vec{g}$ is any body force term, such as gravity.
The fluid equations are coupled by a jump in the stress at the interface,
\begin{equation}
	\left[-p\vec{n}+\mu\left(\nabla\vec{u}+\nabla^T\vec{u}\right)\right]\cdot\vec{n}=\vec{f},
	\label{eq:interfaceJump}
\end{equation}
where $[\;]$ indicates the jump of a quantity (outside minus inside) across the interface and $\vec{f}$ are any forces, such as tension, which act on the interface.

A difficulty in multiphase fluid simulations is the solution of this set of coupled but discontinuous differential equations. Some techniques, such as the
Immersed Interface Method \cite{Lee2003}, augment the discretization of the differential equations to take into account the jumps across the interface. Another technique,
which is explained here, is to model the domain as a ``single" fluid with spatially varying properties \cite{Chang1996, BRACKBILL1992}. The interface condition, Eq. (\ref{eq:interfaceJump}), 
is accounted for by converting the singular force contribution at the interface into a body-force term localized around the interface.

Define the smooth Heaviside, $H_\varepsilon(\phi)$, function as
\begin{equation}
	H_{\varepsilon}(\phi)=
	\begin{cases}
		0 & \phi<-\varepsilon\\
		\tfrac{1}{2}\left[1+\tfrac{\phi}{\varepsilon}+\tfrac{1}{\pi}\sin\left(\tfrac{\pi\phi}{\varepsilon}\right)\right] & |\phi|\leq \varepsilon \\
		1 & \phi>\varepsilon
	\end{cases}
\end{equation}
where $\varepsilon$ is proportional to the grid spacing. 
From the definition of the Heaviside function define the smoothed Dirac-Delta function as $\delta_\varepsilon(\phi)=\partial H_\varepsilon(\phi)/\partial \phi$,
\begin{equation}
	\delta_{\varepsilon}(\phi)=
	\begin{cases}
		0 & |\phi|>\varepsilon\\
		\tfrac{1}{2\varepsilon}\left[1+\cos\left(\tfrac{\pi\phi}{\varepsilon}\right)\right] & |\phi|\leq \varepsilon
	\end{cases}.
\end{equation}
The use of the Heaviside and Dirac functions allows for the calculation of the volume enclosed by a given level set as well as the surface area as a integrals over the domain, 
\begin{align}
	V(\phi)=&\int_\Omega\left(1-H_\varepsilon(\phi)\right)\|\nabla\phi\|dV,\\
	A(\phi)=&\int_\Omega\delta_\varepsilon(\phi)\|\nabla\phi\|\; dV.
\end{align}

Using the Heaviside function the density and viscosity are given by 
$\rho_\varepsilon(\phi)=\rho^-+\left(\rho^+-\rho^-\right)H_\varepsilon(\phi)$ and $\mu_\varepsilon(\phi)=\mu^-+\left(\mu^+-\mu^-\right)H_\varepsilon(\phi)$.
These functions also allow for the transformation of singular interface forces into localized body force terms. For example, let the only
singular interfacial force be from a uniform surface tension, $\vec{f}=\sigma \kappa\vec{n}$, where $\sigma$ is the coefficient of surface tension. 
In conjunction with the smoothed density and viscosity
definitions this results in a single Navier-Stokes equation valid in the entire domain,
\begin{align}
	\rho_\varepsilon(\phi)\dfrac{D\vec{u}}{Dt}=&-\nabla p+\nabla\cdot\left(\mu_\varepsilon(\phi)\left(\nabla\vec{u}+\nabla^T\vec{u}\right)\right)\nonumber\\
	\label{eq:singleNS} 	&-\sigma \delta_\varepsilon(\phi) \kappa\nabla\phi+\vec{g}_\varepsilon(\phi),\\
	\label{eq:singleDiv} \nabla\cdot\vec{u}=&0,
\end{align}
where the body force term has been written as a smooth function of the level-set.
This type of single-fluid formulation for multiphase flow has been used to model bubbles and droplets \cite{Sussman1997,Sussman2012} and vesicles \cite{Kolahdouz2015b,Kolahdouz2015a}.

The results below focus on bubbles and droplets under the influence of surface tension and gravity. Therefore, the interface and body force are restricted
to this particular case. The dimensionless form of the single-fluid Navier-Stokes equations can be obtained by defining a characteristic length $l_0$,
and velocity $u_0$, from which a characteristic time can be obtained, $t_0=l_0/u_0$. The density and viscosity are normalized by the values of the 
outer domain,
\begin{align}
	\rho(\phi)&=\lambda+(1-\lambda)H(\phi), \\
	\mu(\phi)&=\eta+(1-\eta)H(\phi),
\end{align}
where $\lambda=\rho^-/\rho^+$ and $\eta=\mu^-/\mu^+$ are the density and viscosity ratio, respectively, and the $\varepsilon$ has been dropped for clarity.

Define the the Reynolds number as
\begin{equation}
	\textrm{Re}=\dfrac{\rho^+ l_0 u_0}{\mu^+},
\end{equation}
the Weber number as
\begin{equation}
	\textrm{We}=\dfrac{\rho^+ l_0 u_0^2}{\sigma},
\end{equation}
and the Froude number as
\begin{equation}
	\textrm{Fr}=\dfrac{u_0^2}{g_0 l_0},
\end{equation}
where $g_0$ is the strength of gravity pointing in the $\bar{\vec{g}}$ direction. The Navier-Stokes equations now become
\begin{align}
	\rho(\phi)\dfrac{D\vec{u}}{Dt}=&-\nabla p+\dfrac{1}{\textrm{Re}}\nabla\cdot\left(\mu(\phi)\left(\nabla\vec{u}+\nabla^T\vec{u}\right)\right)\nonumber\\
			&-\dfrac{1}{\textrm{We}}\delta(\phi)\kappa\nabla\phi+\dfrac{1}{\textrm{Fr}}(\rho(\phi)-1)\bar{\vec{g}}.
\end{align}
Note that in this particular formulation gravity is written as $(\rho(\phi)-1)\bar{\vec{g}}$ so that uniform acceleration of the entire 
domain does not occur.

\section{A Mass-Preserving Projection Method}
\label{sec:projection-method}
Projection methods are a common technique to solve the Navier-Stokes equations \cite{Brown2001}. In such schemes a tentative velocity field 
is first determined and then the pressure is calculated to enforce the local volume conservation.
In addition to conservation of local volume, Eq. (\ref{eq:singleDiv}), it would be advantageous to explicitly enforce global volume conservation. 
Even if local volume conservation is performed exactly errors in the advection of the interface can introduce unwanted
volume change over the course of a simulation. In this novel projection method both local and global volume conservation are enforced.

The first step is to determine a tentative velocity field at time $t^{n+1}$ in the absence of the pressure and forces.
In this work the semi-Lagrangian formulation is used:
\begin{equation}
	\rho\left(\phi^n\right)\dfrac{\vec{u}^*-\vec{u}^n_d}{\Delta t} = \dfrac{1}{\textrm{Re}}\nabla\cdot\left(\mu\left(\phi^n\right)\left(\nabla\vec{u}^*+\nabla^T\vec{u}^n \right)\right).
	\label{eq:SemiImplicitStep}
\end{equation}
The departure velocity, $\vec{u}^n_d$, for a grid point $\vec{x}$ is determined by tracing characteristics backward in time,
\begin{align}
	\vec{x}_d &= \vec{x} - \Delta t\vec{u}^n\left(\vec{x}\right), \\
	\vec{u}_d^n &= \vec{P}_u\left(\vec{x}_d,t^n\right),
\end{align}
where $\vec{P}_u$ is an interpolant of the velocity field at time $t^n$. Additional information regarding the semi-Lagrangian method for Navier-Stokes equations
is given in Ref \cite{Xiu2001}.

The next step of a standard projection method would be to determine the pressure to satisfy local volume conservation, Eq. (\ref{eq:singleDiv}).
In this work global conservation is also considered. Global volume conservation can be enforced by looking at the rate
at which the enclosed volume, $V$, will change over time \cite{Kolahdouz2015b,Laadhari2012}
\begin{equation}
	\int_\Gamma\vec{n}\cdot\vec{u}^{n+1} dA = \dfrac{dV}{dt}
	\label{eq:GlobalConservation}
\end{equation}
for an outward facing unit normal, $\vec{n}$, to the interface $\Gamma$.
Unfortunately, the pressure field alone can not satisfy both constraints. Therefore the pressure is split into a constant and spatially-varying component:
\begin{equation}
	p = \tilde{p} + (1-H(\phi))p_0.
\end{equation}
The constant pressure component, $p_0$, only has a contribution to the overall pressure field in the region given by $\phi<0$ and can be thought of enforcing 
global volume conservation. The spatially-varying component, $\tilde{p}$, takes the place of the standard pressure and will be used to enforce 
local volume conservation. Expanding the pressure component results in 
\begin{equation}
	-\nabla p = -\nabla(\tilde{p}+(1-H(\phi^n))p_0) = -\nabla\tilde{p} + \delta(\phi^n)p_0\nabla\phi^n,
\end{equation}
which demonstrates that $p_0$ only has a contribution near the interface.

Using this definition of pressure the projection step is now
\begin{align}
	\dfrac{\vec{u}^{n+1}-\vec{u}^*}{\Delta t} =& -\dfrac{1}{\rho(\phi^n)}\nabla\tilde{p}+\dfrac{\delta(\phi^n)p_0\nabla\phi^n}{\langle\rho\rangle}\nonumber\\
				&-\dfrac{\delta(\phi^n)\kappa\nabla\phi}{\langle\rho\rangle\textrm{We}}+\dfrac{\rho(\phi^n)-1}{\rho(\phi^n)\textrm{Fr}}\bar{\vec{g}}.
	\label{eq:projection}
\end{align}
For quantities localized around the interface the average density is used, $\langle\rho\rangle=(\rho^+ + \rho^-)/2$.

To determine $\tilde{p}$ and $p_0$ both conservation conditions are applied to the projection step, Eq. (\ref{eq:projection}).
Applying local conservation, Eq. (\ref{eq:singleDiv}), results in
\begin{align}
	\nabla\cdot\left(\dfrac{1}{\rho(\phi^n)}\nabla\tilde{p}\right) - \dfrac{p_0}{\langle\rho\rangle}\nabla\cdot\left(\delta(\phi^n)\nabla\phi\right) = \nonumber\\
	\nabla\cdot\left(\dfrac{\vec{u}^*}{\Delta t}+\dfrac{\delta(\phi^n)\kappa\nabla\phi}{\langle\rho\rangle\textrm{We}}-\dfrac{\rho(\phi^n)-1}{\rho(\phi^n)\textrm{Fr}}\bar{\vec{g}}\right).	
	\label{eq:localSolve}
\end{align}
Global conservation can be enforced by combining Eqs. (\ref{eq:GlobalConservation}) and (\ref{eq:projection}) and also
requiring that the time-evolution of the volume be such that any previous errors are corrected,
\begin{align}
	\int_\Gamma \dfrac{1}{\rho(\phi^n)}\vec{n}\cdot\nabla\tilde{p}\;dA-\dfrac{p_0}{\langle\rho\rangle}\int_\Gamma\delta(\phi^n)\|\nabla\phi\|\;dA =\nonumber\\
	\int_\Gamma\left(\dfrac{\vec{n}\cdot\vec{u}^*}{\Delta t} +\dfrac{\delta(\phi^n)\kappa\|\nabla\phi\|}{\langle\rho\rangle\textrm{We}}
		-\dfrac{\rho(\phi^n)-1}{\rho(\phi^n)\textrm{Fr}}\vec{n}\cdot\bar{\vec{g}}\right)dA\nonumber\\
		-\dfrac{1}{\Delta t}\dfrac{V^0-V^n}{\Delta t}		
	\label{eq:globalSolve}
\end{align}
where $V^0$ is the initial enclosed volume. 

Let $\tilde{\vec{p}}$ represent the vector holding the the values of $\tilde{p}$ over the entire discretized domain. 
The surface integrals in Eq. (\ref{eq:globalSolve}) can be written as a summation over a discretized domain, $\int_\Gamma f dA\approx \sum\delta_{i,j,k}f_{i,j,k}dV$
where grid points are given by $\vec{x}_{i,j,k}$, $\delta_{i,j,k}=\delta(\phi(\vec{x}_{i,j,k}))$, $f_{i,j,k}$ is the function value at the grid points, and $dV$ is the 
volume of each cell surrounding a grid point.
It is then possible to define the following 
linear operators:
\begin{align}
	\label{eq:linear1} \nabla\cdot\left(\dfrac{1}{\rho(\phi^n)}\nabla\tilde{p}\right) \approx&\;\vec{L}\tilde{\vec{p}}, \\
	\label{eq:linear2} \dfrac{p_0}{\langle\rho\rangle}\nabla\cdot\left(\delta(\phi^n)\nabla\phi\right) \approx&\;p_0 \vec{l},\\
	\label{eq:linear3} \int_\Gamma \dfrac{1}{\rho(\phi^n)}\vec{n}\cdot\nabla\tilde{p}\;dA \approx&\;\vec{s}^T\tilde{\vec{p}}\\
	\label{eq:linear4} \dfrac{p_0}{\langle\rho\rangle}\int_\Gamma\delta(\phi^n)\|\nabla\phi\|\;dA \approx&\; p_0 a.
\end{align}
The two constraint equations, Eqs. (\ref{eq:localSolve}) and (\ref{eq:globalSolve}), can now be written in block matrix-vector form,
\begin{equation}
	\begin{bmatrix}
		\vec{L} & \vec{l} \\
		\vec{s}^T & a
	\end{bmatrix} \;
	\begin{bmatrix}
		\tilde{\vec{p}} \\
		p_0
	\end{bmatrix} = 
	\begin{bmatrix}
		\vec{d} \\
		e
	\end{bmatrix},
\end{equation}
where $\vec{d}$ is the discretization of the right-hand-side of Eq. (\ref{eq:localSolve}) while $e$ is the evaluation of the integral on the right-hand-side of Eq. (\ref{eq:globalSolve}).
This particular block matrix form can be solved efficiently through the use of a Schur decomposition,
\begin{equation}
	\begin{bmatrix} 
		\tilde{\vec{p}}\\ 
		p_0 
	\end{bmatrix} = 
	\begin{bmatrix} 
		\vec{I} & 0 \\ 
		-a^{-1}\vec{s}^T & 1
	\end{bmatrix} 
	\begin{bmatrix} 
		\vec{S}^{-1} & 0 \\ 
		0 & a^{-1}
	\end{bmatrix} 
	\begin{bmatrix} 
		\vec{I} & -a^{-1}\vec{l} \\ 
		0 & 1
	\end{bmatrix} 
	\begin{bmatrix}
		\vec{d}\\
		e
	\end{bmatrix}.
\end{equation}
The Schur complement is a rank-1 update on the matrix $\vec{L}$:
\begin{equation}
	\vec{S}=\vec{L}-\dfrac{1}{a}\vec{l}\vec{s}^T,
\end{equation}
which has an inverse given by the Sherman-Morrison formula,
\begin{equation}
	\vec{S}^{-1}=\vec{L}^{-1}+\dfrac{\vec{L}^{-1}\vec{l}\vec{s}^T\vec{L}^{-1}}{a-\vec{s}^T\vec{L}^{-1}\vec{l}}.
\end{equation}
So long as $a-\vec{s}^T\vec{L}^{-1}\vec{l}$ is non-zero the inverse is defined. This condition is not proven here, but a check is performed
during all simulations and has never been violated. Note that as the method as described does not depend on any particular spatial
discretization. The particular discretization used in the numerical experiments below will be discussed in Sec. \ref{sec:projectionDiscrete}.

\section{Numerical Implementation}
\label{sec:numerics}
In this section the particular numerical implementation used in the numerical experiments of Sec. \ref{sec:experiments} is briefly discussed. 
In particular the description and advection of the interface and the discretization of the projection method is presented.
\subsection{Interface Description and Advection}

The interface separating the multiple fluid phases is described here using a gradient-augmented level set method \cite{Nave2010,Seibold2012}. In this extension 
of the level set method the gradient field of the level set is explicitly tracked in addition to the level set function. This allows for the accurate 
determination of interfacial quantities such as curvature \cite{Nave2010, Seibold2012}. In particular the results shown below use a recent semi-implicit
extension of the original level set Jet scheme \cite{Velmurugan2015}. 

Consider the implicit tracking of an interface by way of a level set function: $\Gamma(t)=\{\vec{x}:\phi(\vec{x},t)=0\}$. In addition to the level set
also track the gradient field of the level set, $\vec{\psi}=\nabla\phi$. This allows for the determination of Hermite interpolants using only 
information local to a cell on a Cartesian grid \cite{Nave2010}. Unfortunately, the original Jet scheme does not work well for stiff
advection equations, such as those involving surface tension, and the recent SemiJet method was developed in response \cite{Velmurugan2015}. 

The SemiJet method begins by performing a semi-implicit, semi-Lagrangian update on the level set field,
\begin{equation}
	\dfrac{\phi^{n+1}-\phi^n_d}{\Delta t}=\beta\nabla^2 \phi^{n+1}-\beta\nabla^2 \phi^n,
	\label{eq:semijet-phi}
\end{equation}
where $\phi^n_d$ is the level set value at the departure location and $\beta=0.5$ is a constant. \

To update the gradient field note that the smoothing operation can be captured using a smoothing operator,
\begin{equation}
	S_\phi=\beta\nabla^2 \phi^{n+1}-\beta\nabla^2 \phi^n,
	\label{eq:semijet-source}
\end{equation}
where $\phi^{n+1}$ and $\phi^n$ are both known at grid points after solving Eq. (\ref{eq:semijet-phi}). At each grid point define a sub-grid
of spacing $\epsilon\ll h$, where $h$ is the grid spacing:
\begin{equation}
	\vec{x}^{\vec{q}}=\vec{x}+\vec{q}\epsilon \textrm{ for } \vec{q}\in\{-1,1\}^p,
\end{equation}
with $p=2$ in two-dimensions and $p=3$ in three-dimensions. This results in a sub-grid of either 4 or 8 points, depending on the dimension of the simulation.
On each of these sub-grid points an updated level-set value is obtained using an explicit semi-Lagrangian update with the smoothing source term 
given in Eq. (\ref{eq:semijet-source}),
\begin{equation}
	\dfrac{\phi^{\vec{q}}-\phi^{\vec{q}}_d}{\Delta t}=S_{\phi}\left(\vec{x}^{\vec{q}}\right).
	\label{eq:semijet-subgrid}
\end{equation}
Once the level set values on the subgrid are calculated updated gradient values at grid points can be obtained using finite difference approximations. For example,
in two-dimensions the gradient in the $x$-direction at a grid is given by $\left(\phi^{(1,1)}-\phi^{(-1,1)}+\phi^{(1,-1)}-\phi^{(-1,-1)}\right)/4\epsilon$.

Curvature values are calculated at grid points using
\begin{equation}
	\kappa = \frac{\phi_{xx}\phi_y^2+\phi_{yy}\phi_x^2-2\phi_{xy}\phi_x\phi_y}{\left(\phi_x^2+\phi_y^2\right)^{3/2}},	
\end{equation}
in two-dimensions with a similar expression for three-dimensions. 
The values used to calculate the curvature were calculated by averaging derivatives. For example, when computing $\phi_x$, $\phi_{xx}$ and $\phi_{xy}$ 
for use in the curvature calculation the following averaging would be used:
\begin{align}
	\phi_x &= \tfrac{1}{2}\left(\psi^x + D_x\phi\right),\\
	\phi_{xx} &= \tfrac{1}{2}\left(D_x\psi^x + D_{xx}\phi\right),\\
	\phi_{xy} &= \tfrac{1}{3}\left(D_y\psi^x + D_x\psi^y + D_{xy}\phi\right),
\end{align}
where $D_x$, $D_y$, $D_{xx}$ and $D_{xy}$ are the finite difference approximations for the derivatives and $\psi^x$ and $\psi^y$ are the components of the 
gradient field tracked by the Jet. The curvature was then extended away from the interface in the normal direction.
Complete information about the SemiJet method and curvature calculation can be found in Ref. \cite{Velmurugan2015}.

\subsection{Discretization of the Projection Method}
\label{sec:projectionDiscrete}
The Navier-Stokes equations and projection method are discretized on a collocated Cartesian mesh using standard finite difference approximations. The variable coefficient
Poisson equation, Eq. (\ref{eq:linear1}), is discretized using compact second-order finite difference approximations. In one-dimension this is written as
\begin{equation}
	\vec{L}\tilde{p}_{i}=\dfrac{\rho^{-1}_{i+1/2}\left(\tilde{p}_{i+1}-\tilde{p}_{i}\right)-\rho^{-1}_{i-1/2}\left(\tilde{p}_{i}-\tilde{p}_{i-1}\right)}{h^2}
\end{equation}
where the inverse density at half-grid points is obtained using harmonic averaging,
\begin{equation}
	\rho^{-1}_{i+1/2}=\dfrac{\rho_{i}+\rho_{i+1}}{2\rho_{i}\rho_{i+1}}.
\end{equation}
It was found that the harmonic averaging improved the stability for large density variations. Similar expressions hold in other dimensions. 
The remaining discretizations in Eqs. (\ref{eq:linear2})-(\ref{eq:linear4}) were also performed using centered second-order finite difference approximations.

Note that the method described here is an approximate pressure discretization. It is well known that this particular discretization does not suffer from
the pressure decoupling effect which can occur from a discrete discretization on a collocated mesh \cite{PERIC1988,Rauwoens2007}. On the other hand the fact that
the divergence-free velocity field condition is not discretely enforced means that small mass errors will accumulate over time. An approximate discretization was
purposefully chosen to demonstrate the ability of the proposed method to conserve mass even if the underlying method is not discretely mass conserving.

\section{Numerical Experiments}
\label{sec:experiments}
The mass-preserving projection method described above is tested using four different multiphase fluid systems. These are: 1) a static bubble, 2) a rising bubble, 3) a 
fluid thread undergoing a Plateau-Rayleigh instability, and 4) a droplet in shear flow. In all cases the mass-preserving method, denoted as the conserving scheme, is compared
to a simulation without the method described above. This non-conserving scheme is computed by using the same discretization as the conserving scheme with 
the pressure calculated as the solution to Eq. (\ref{eq:localSolve}) with $p_0=0$.
All of the results will be three-dimensional results using uniform grid spacing, although some figures will show two-dimensional slices instead of three-dimensional figures.

\subsection{A Static Bubble}
\begin{figure}[!ht]
	\begin{center}
		\subfigure[t=0]{
			\includegraphics[width=1.0in]{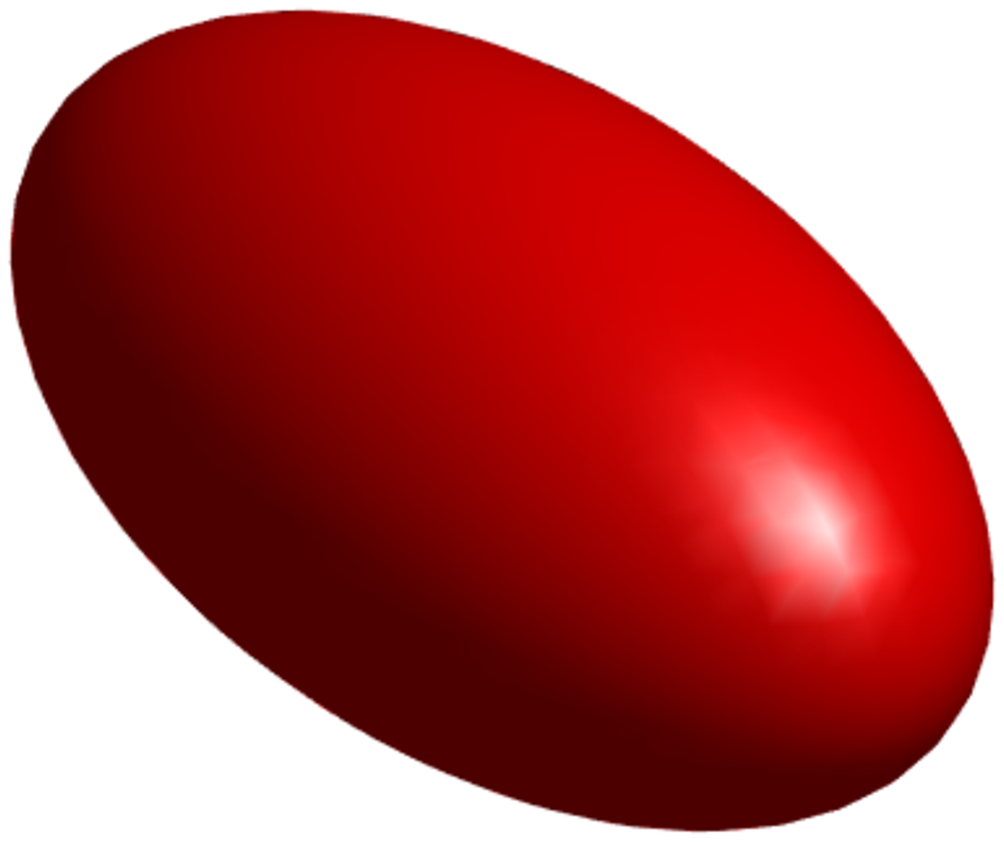}		
		}
		\subfigure[t=0.5]{
			\includegraphics[width=1.0in]{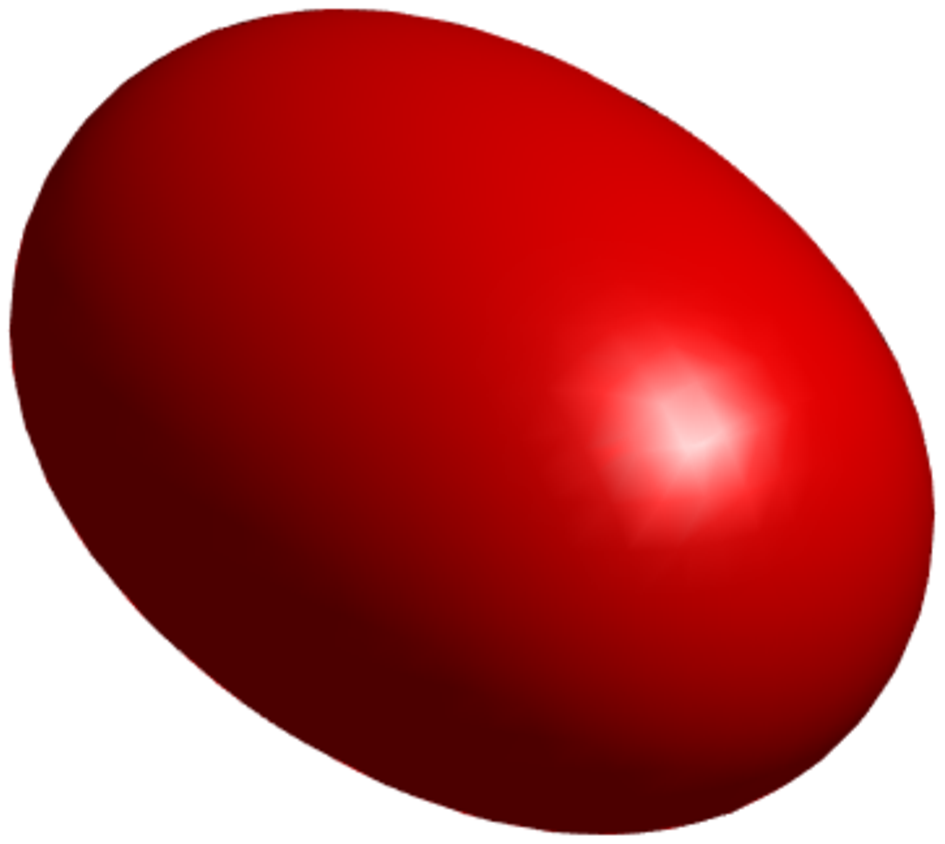}		
		}
		\subfigure[t=1]{
			\includegraphics[width=1.0in]{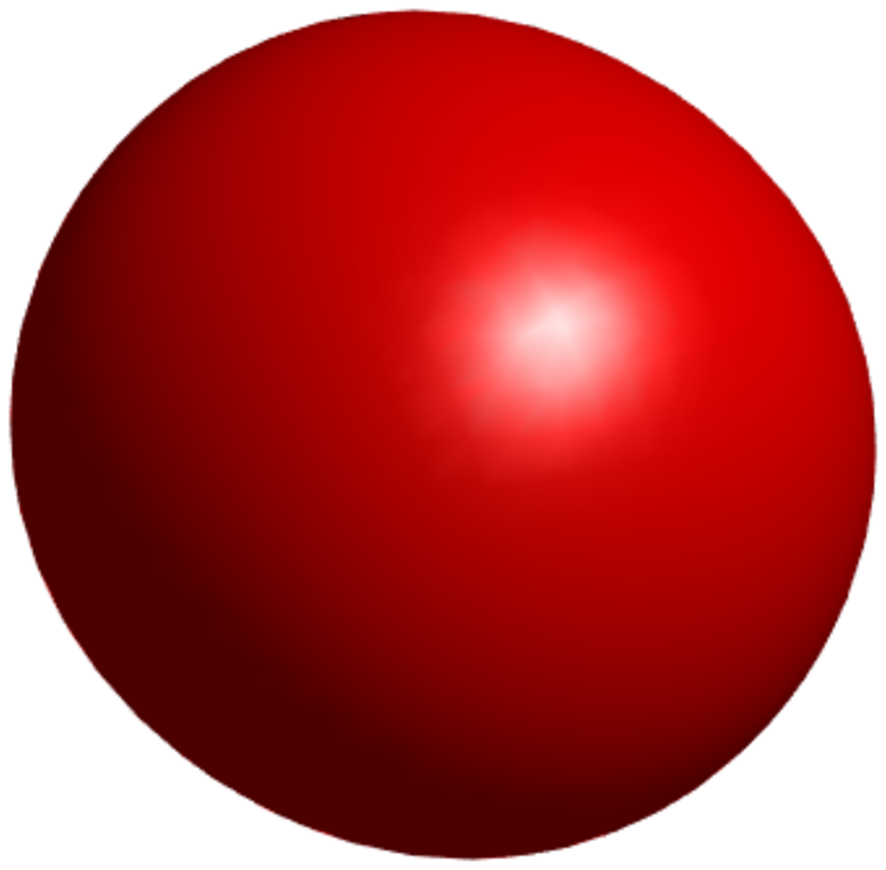}		
		}\\
		\subfigure[t=2]{
			\includegraphics[width=1.0in]{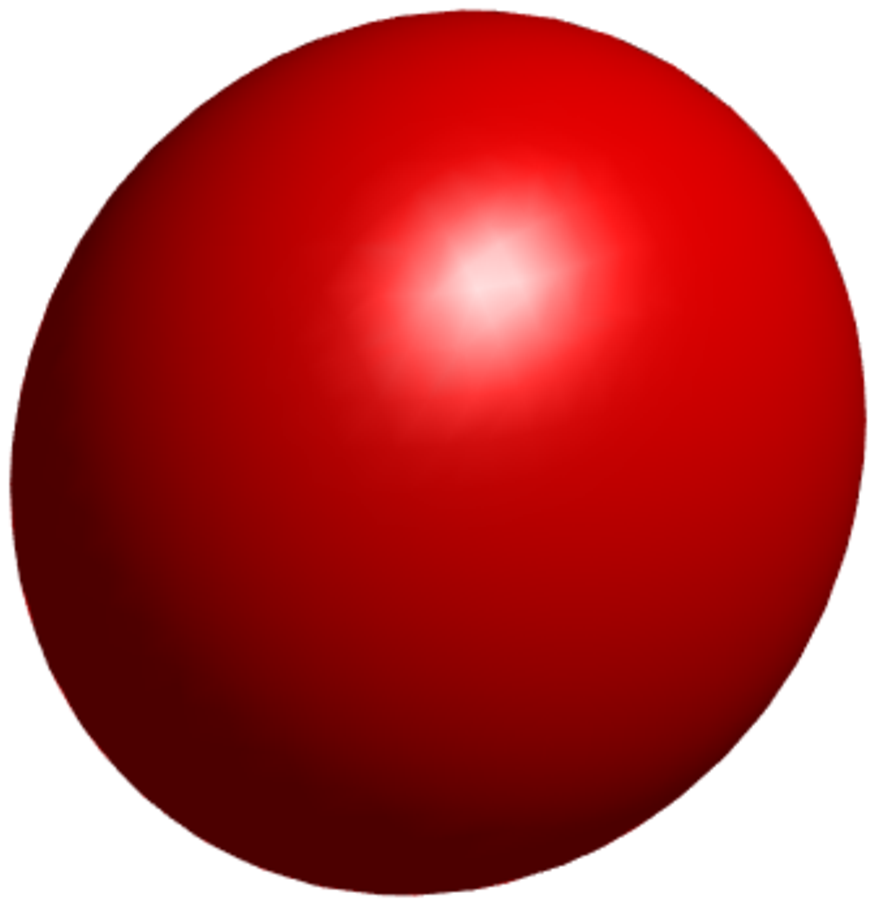}		
		}		
		\subfigure[t=10]{
			\includegraphics[width=1.0in]{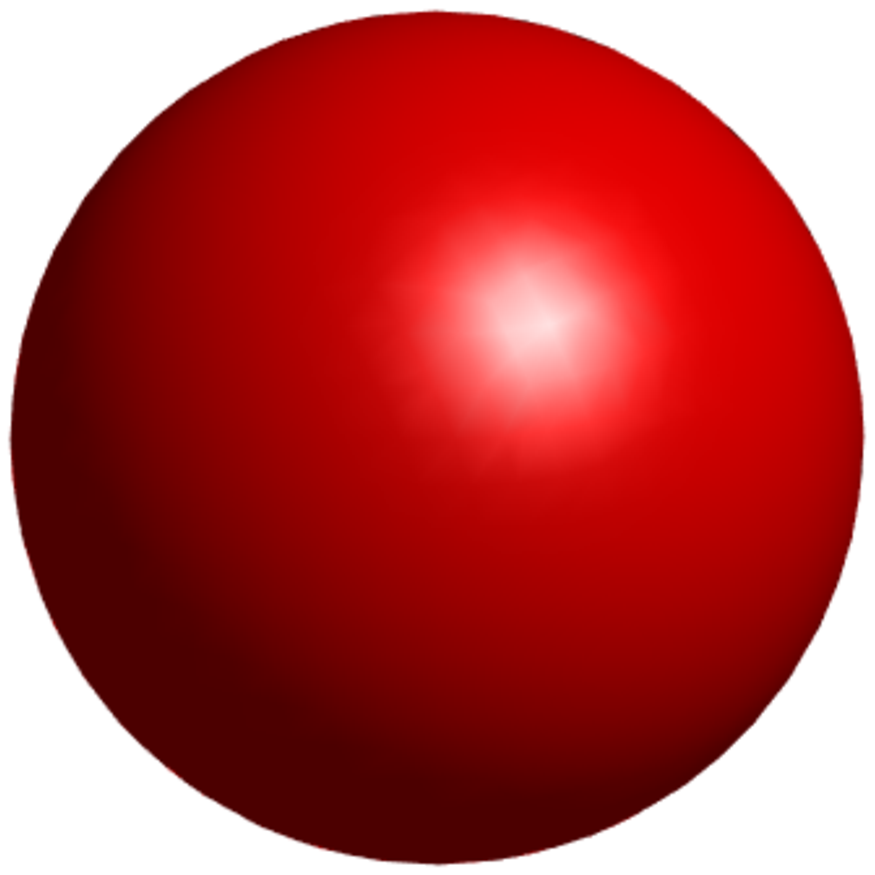}		
		}		
		\subfigure[t=20]{
			\includegraphics[width=1.0in]{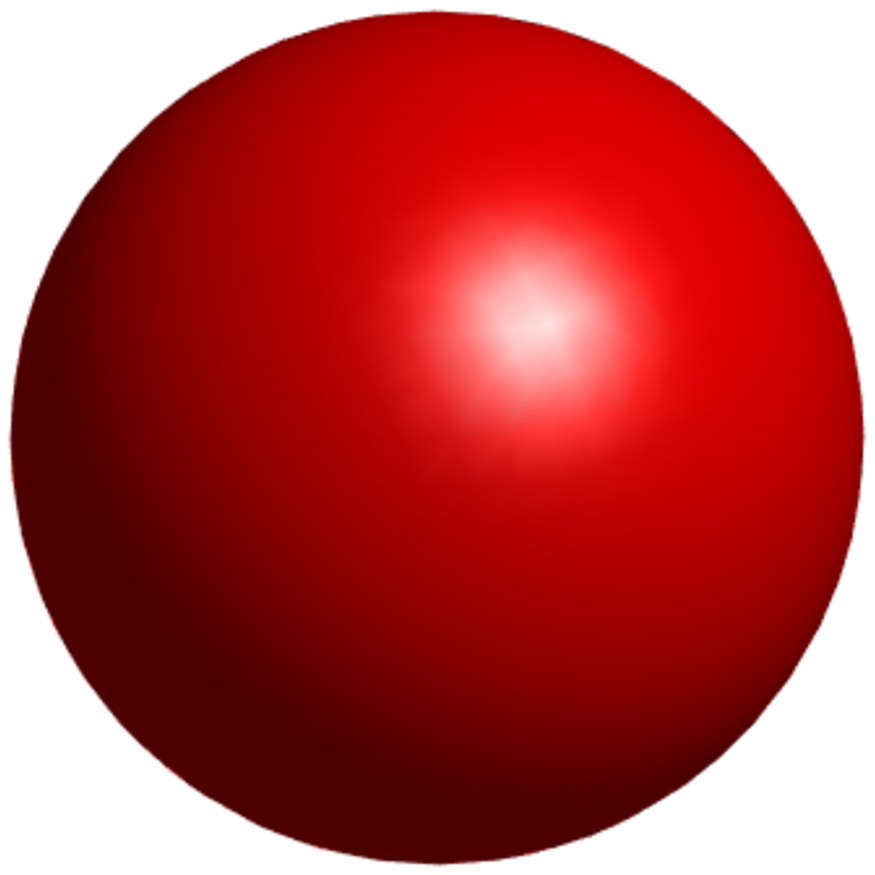}		
		}
		\caption{Evolution of a bubble under surface tension with $\textrm{Re}=10$, $\textrm{We}=1$, and $\textrm{Fr}=\infty$. The density ratio is $\lambda=0.001$ while
		the viscosity ratio is $\eta=0.01$. The final sphere radius is 1.261.}	
		\label{fig:StaticBubbleEvolution}	
	\end{center}
\end{figure}

The first numerical experiment is the evolution of a droplet in the absence of gravity, Fig. \ref{fig:StaticBubbleEvolution}. 
The viscosity ratio is set to $\eta=0.01$ while the density ratio
is $\lambda=0.001$. The Reynolds number is 10 and the Weber number is 1. Gravity is ignored, which is equivalent to setting $\textrm{Fr}=\infty$. 
The initial shape is an ellipsoidal interface with an axis of 1 in the $x-$ and $z-$directions and 2 in the $y-$direction.
The computational domain is a cube given by $[-4,4]^3$ and a $65^3$ grid, resulting in a grid spacing of $h=0.125$. The time step is $\Delta t=0.01$.

Under the influence of surface tension the bubble evolves into a sphere with the same volume. It is therefore expected that 
the final shape will be a sphere with a radius of 1.26. The simulated final radius is 1.261, very close to the theoretical value.

\begin{figure}[!ht]
	\begin{center}
		\subfigure[Spatially Varying Pressure Component: $\tilde{p}$]{
			\includegraphics[width=2.75in]{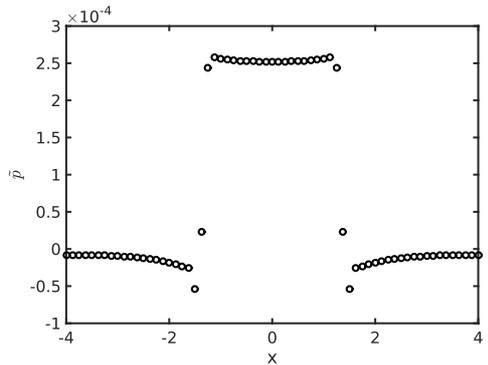}		
		}
		\subfigure[Total Pressure: $p=\tilde{p}+(1-H(\phi))p_0$]{
			\includegraphics[width=2.75in]{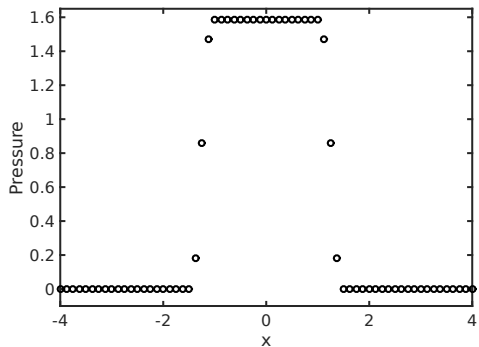}
		}		
		\caption{The pressure along the $x$-axis with $y=0$ and $z=0$ at a time of $t=20$ for the static bubble shown in Fig. \ref{fig:StaticBubbleEvolution}.
		The spatially varying pressure component, $\tilde{p}$, is nearly zero while the pressure constant is $p_0=1.584$. This is very near to the predicted static pressure of 1.587.}
		\label{fig:StaticBubblePressure}
	\end{center}
\end{figure}

The distribution of the spatially-varying pressure component, $\tilde{p}$, 
and the total pressure $p$ along the $x-$axis for $y=0$ and $z=0$ at a time of $t=20$ 
are shown in Fig. \ref{fig:StaticBubblePressure}. Results are similar in the other directions.
It is clear that at steady-state the majority of the pressure is due 
to $p_0$, which has a value of $p_0=1.584$. 
According to the Laplace-Young condition it is also expected that the difference between the inner and outer
pressure should be $p^- -p^+=2\kappa/\textrm{We}$. Using an analytic final radius of 1.26 and the simulation
parameters this results in an expected pressure jump of 1.587, which matches well with the simulated final pressure.

\begin{figure}[!ht]
	\begin{center}
		\subfigure[Volume]{
			\includegraphics[width=2.75in]{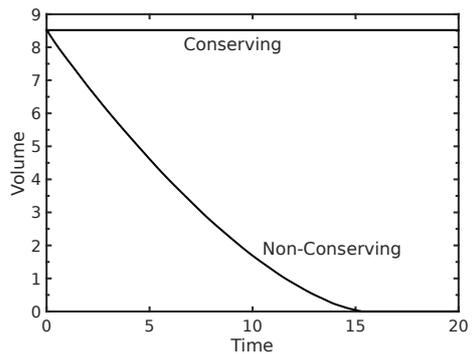}		
		}
		\subfigure[Volume Error]{
			\includegraphics[width=2.75in]{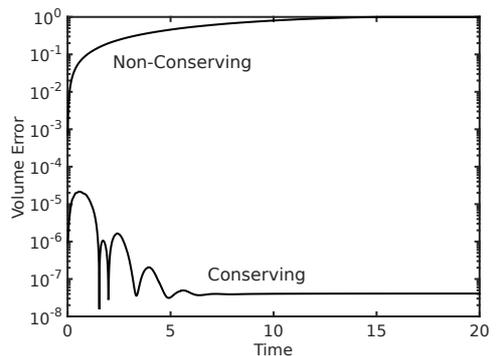}
		}		
		\caption{Comparison between volume preservation in the conserving and non-conserving schemes for a static bubble with the initial condition shown in Fig. \ref{fig:StaticBubbleEvolution}.
			After $t=15$ the bubble has completely disappeared in the non-conserving simulation.}
		\label{fig:StaticBubbleVolume}
	\end{center}
\end{figure}

A comparison between the mass conserving and non-mass conserving simulations is given in Fig. \ref{fig:StaticBubbleVolume}. Both the volume and volume-error over the 
course of the simulation are presented. The volume error is computed as $(V(t)-V(0))/V(0)$ where $V(t)$ is the instantaneous volume and $V(0)$ is the initial volume.
The use of an approximate projection method, combined with the non-conservation properties of the level set method, results in excessive mass loss in the non-conserving
simulation. This mass loss results in the complete disappearance of the bubble in the non-conserving simulation, while the conserving scheme has a maximum volume 
error of approximately $2\times 10^{-5}$.

\subsection{A Rising Bubble}
The next numerical experiment is that of a rising bubble. 
The computational domain is a cube given by $[-5,5]^3$ using a $101^3$ grid, giving a grid spacing of 0.1. The time step is chosen to be $0.01$.
The initial bubble is a sphere centered at the origin with a radius of 1.

To allow for the long-time evolution of the bubble to be modeled after every time step the level set Jet and complete fluid field are re-centered in the computational
domain using a cubic interpolation procedure. Using the values for the spherical cap Bubble A in Table 1 of Hnat and Buckmaster \cite{HNAT1976} 
the density ratio is set to $\lambda=0.001142$ while the viscosity ratio is $\eta=0.008474$. Based on the dimension of the experimental bubble, the final
rise velocity, and the given surface tension the dimensionless parameters are $\textrm{Re}=9.694$, $\textrm{We}=7.6375$ and $\textrm{Fr}=0.7754$.
\begin{figure}[!ht]
	\begin{center}
		\subfigure[t=0]{
			\includegraphics[width=1.0in]{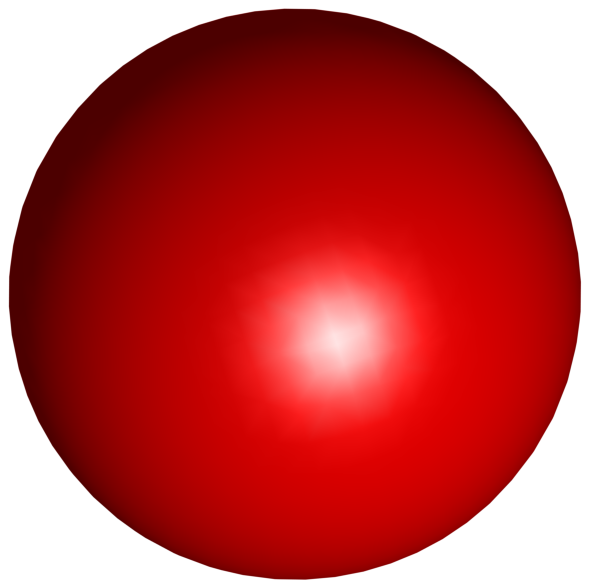}		
		}
		\subfigure[t=1]{
			\includegraphics[width=1.0in]{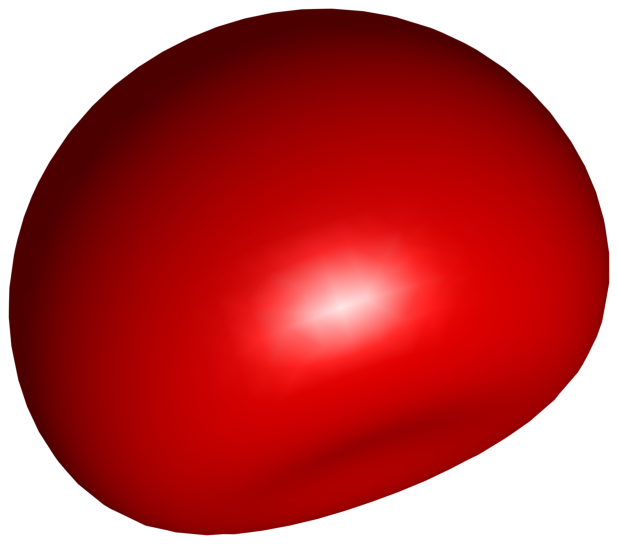}
		}
		\subfigure[t=2]{
			\includegraphics[width=1.0in]{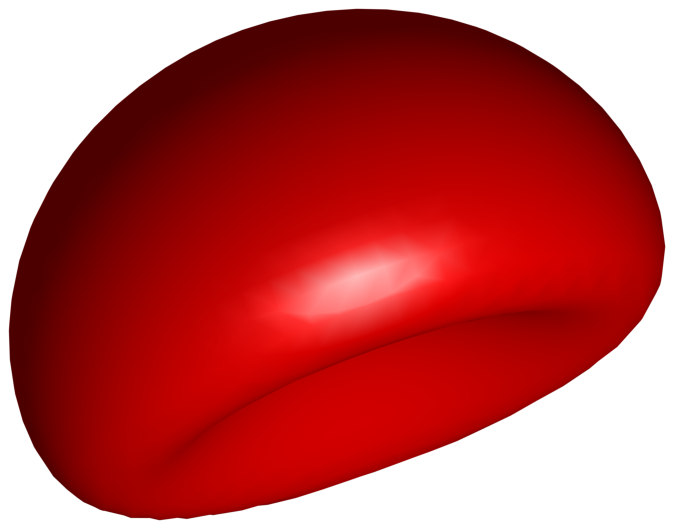}
		}\\		
		\subfigure[t=4]{
			\includegraphics[width=1.0in]{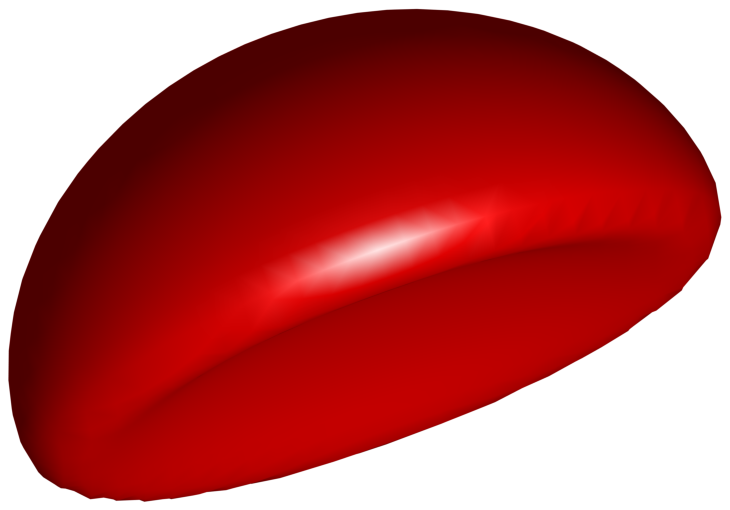}
		}
		\subfigure[t=10]{
			\includegraphics[width=1.0in]{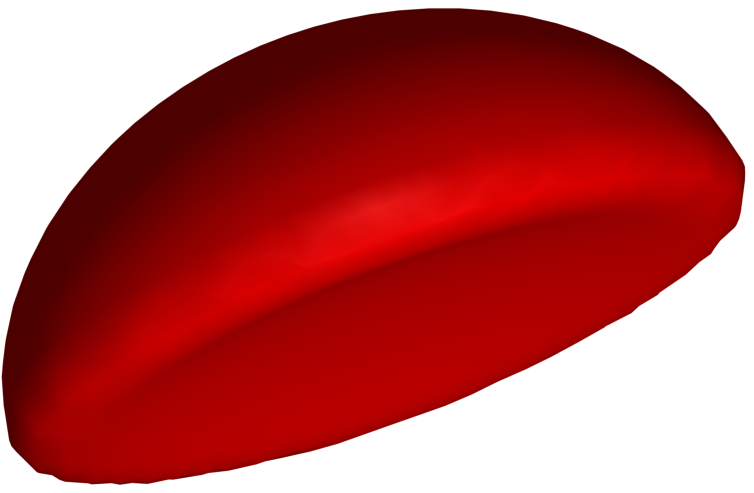}
		}		
		\subfigure[t=20]{
			\includegraphics[width=1.0in]{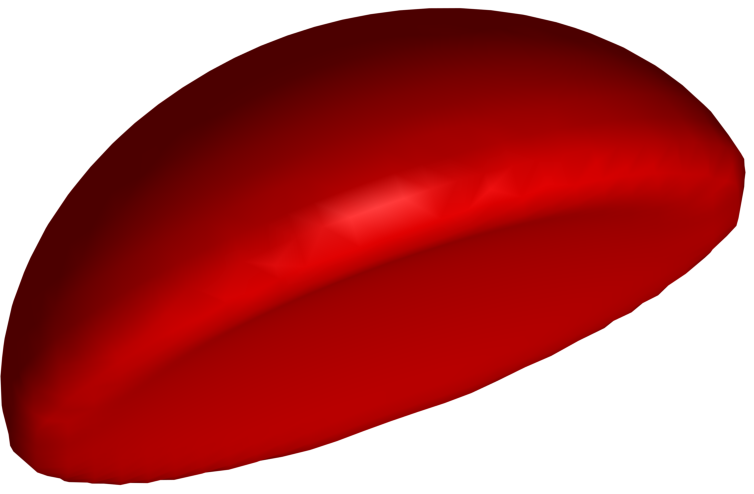}
		}
		\caption{Evolution of a three-dimensional bubble with density ratio of $\lambda=0.001142$ and viscosity ratio of $\eta=0.008474$ with $\textrm{Re}=9.694$,
		$\textrm{We}=7.6375$ and $\textrm{Fr}=0.7754$. These properties correspond to Bubble A in Table 1 in Hnat and Buckmaster \cite{HNAT1976}. Experimental results show the aspect 
		ratio should be 2.7 while the simulation results in an aspect ratio of 2.8.}
		\label{fig:BubbleRiseEvolution3D}	
	\end{center}
\end{figure}

\begin{figure}[!ht]
	\begin{center}
		\subfigure[t=0]{
			\includegraphics[width=1.0in]{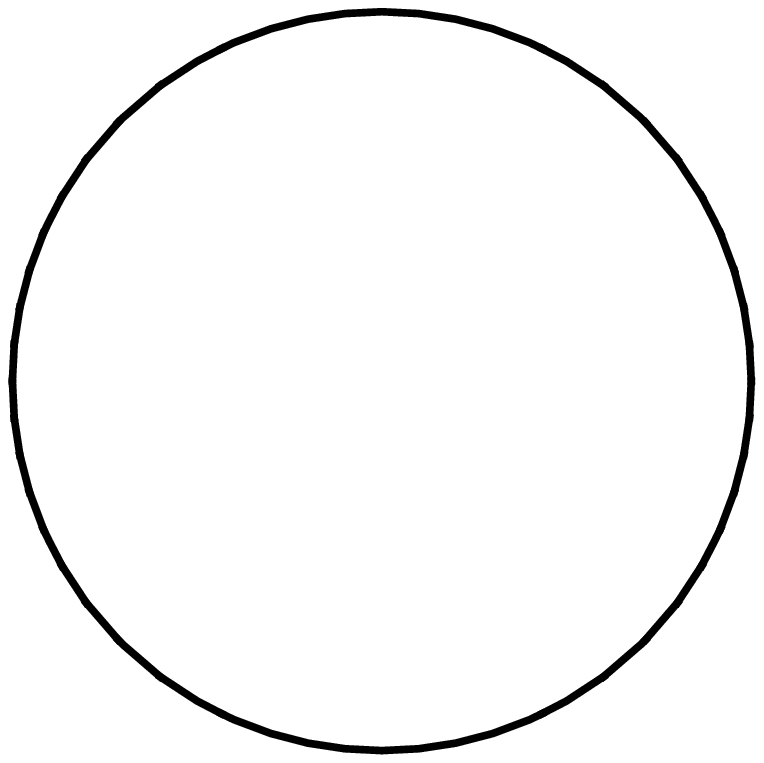}		
		}
		\subfigure[t=1]{
			\includegraphics[width=1.0in]{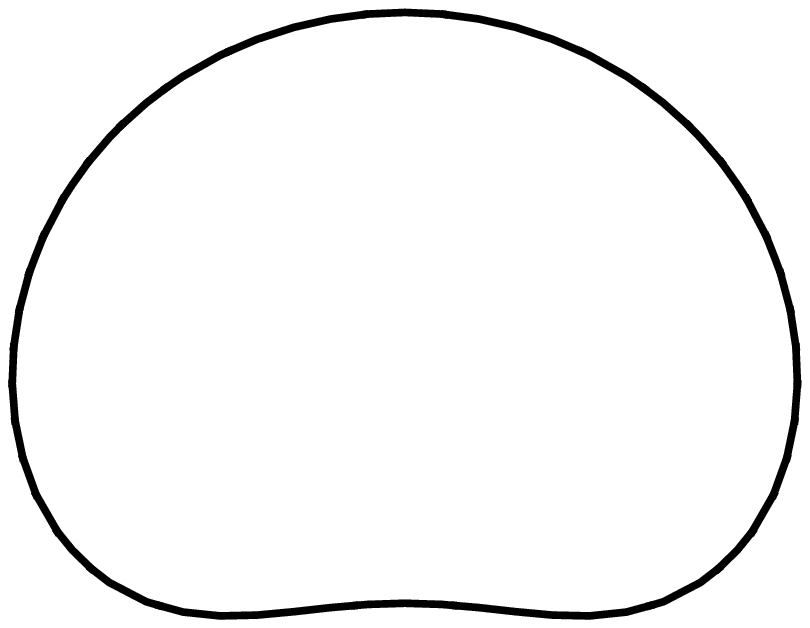}
		}
		\subfigure[t=2]{
			\includegraphics[width=1.0in]{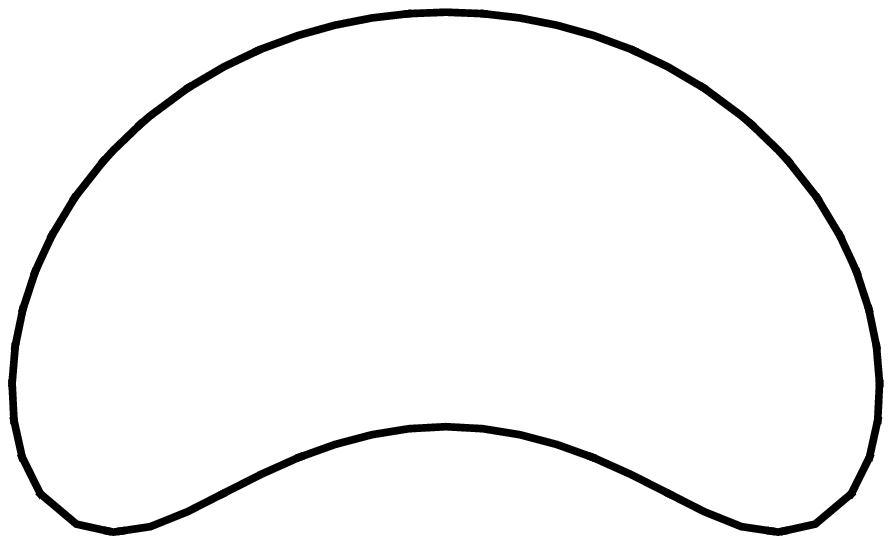}
		}\\		
		\subfigure[t=4]{
			\includegraphics[width=1.0in]{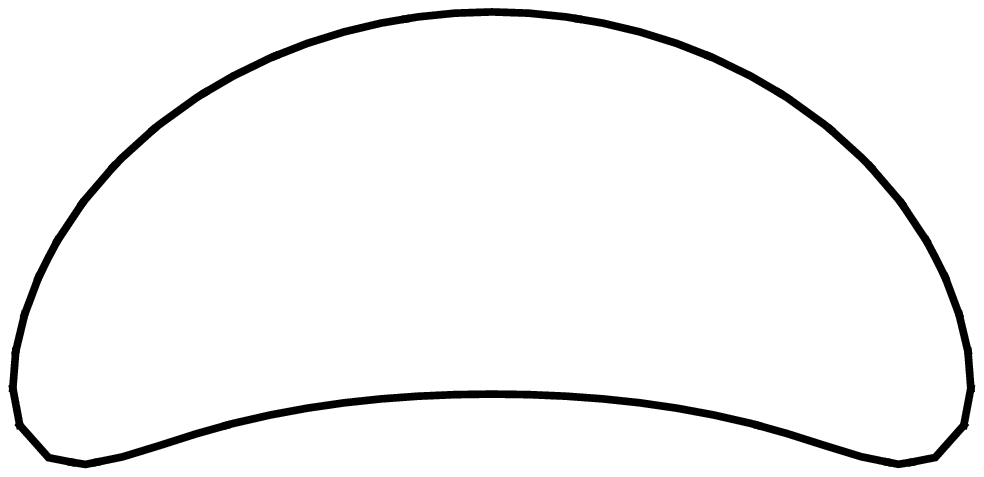}
		}
		\subfigure[t=10]{
			\includegraphics[width=1.0in]{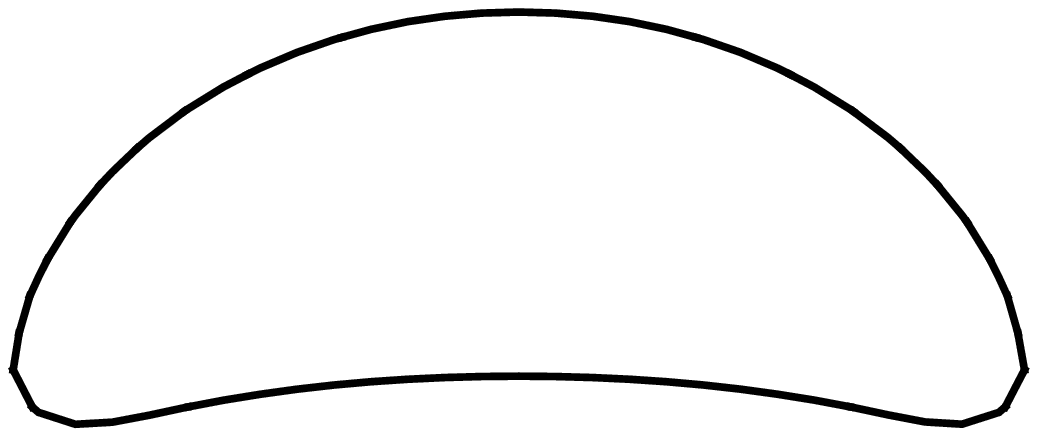}
		}		
		\subfigure[t=20]{
			\includegraphics[width=1.0in]{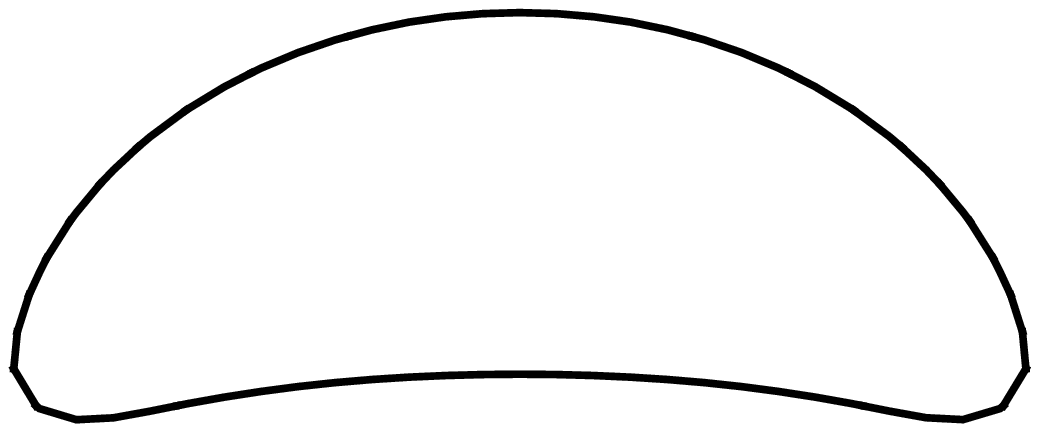}
		}
		\caption{Evolution of the $x-y$ plane of a three-dimensional bubble for the result shown in Fig. \ref{fig:BubbleRiseEvolution3D}.}
		\label{fig:BubbleRiseEvolution}	
	\end{center}
\end{figure}

\begin{figure}[!ht]
	\begin{center}		
		\includegraphics[width=2.75in]{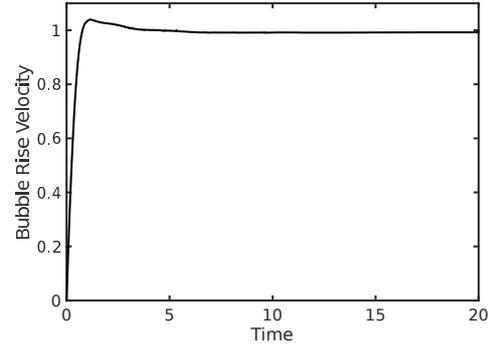}				
		\caption{The rise velocity of the center of mass for the three-dimensional bubble in Figs. \ref{fig:BubbleRiseEvolution3D} and \ref{fig:BubbleRiseEvolution}. The experimental results of Hnat and
		Buckmaster \cite{HNAT1976} indicate the final dimensionless rise velocity should be 1. The simulation results in a rise velocity of 0.9922.} 
		\label{fig:BubblRiseVelocity}
	\end{center}
\end{figure}

As the bubble rises it evolves into the expected spherical cap, see Fig. \ref{fig:BubbleRiseEvolution3D} for a three-dimensional figure and Fig. \ref{fig:BubbleRiseEvolution}
for the cross-section in the $x-y$ plane. The experimental results indicate that the aspect ratio of the bubble at steady-state should be 2.7, which compares to a value of 2.8 from
the simulation. With the given parameter set it is expected that the dimensionless rise velocity should be equal to one. The rise velocity over the course of the 
simulation is given in Fig. \ref{fig:BubblRiseVelocity}, with a steady-state rise velocity of 0.9922.

\begin{figure}[!ht]	
	\begin{center}
		\subfigure[Volume]{
			\includegraphics[width=2.75in]{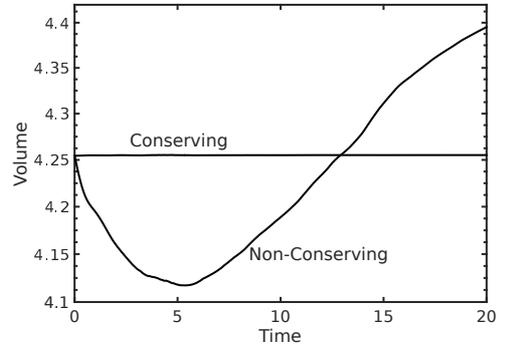}		
		}
		\subfigure[Volume Error]{
			\includegraphics[width=2.75in]{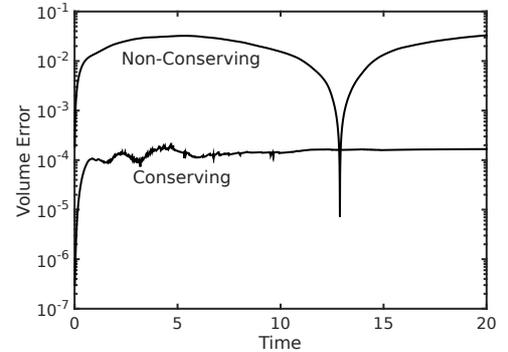}
		}		
		\caption{Comparison between volume preservation in the conserving and non-conserving schemes for a rising bubble with the initial condition shown 
			in Figs. \ref{fig:BubbleRiseEvolution3D} and \ref{fig:BubbleRiseEvolution}.
			The volume gain will continue in the non-conserving scheme while the volume will remain constant in the conserving scheme.}
		\label{fig:RisingBubbleVolume}
	\end{center}	
\end{figure}

The evolution of the volume over the course of the simulation is presented in Fig. \ref{fig:RisingBubbleVolume}. Note that due to discretization errors
the initial volume is not equal to $4\pi/3\approx 4.19$ but is calculated numerically as 4.252.
As with the static bubble example both the volume and the 
volume error are given. Over the course of the simulation the volume for the non-conserving scheme varies from a minimum of 4.13 to a maximum of 4.39, while the
conserving scheme only varies slightly. It should also be obvious that if the simulation was allowed to continue volume errors in the non-conserving scheme 
would continue to grow, which is not true for the conserving scheme.

\subsection{Plateau-Rayleigh Instability}
As a third test case consider the evolution of a fluid thread. It is well known that if the shape has an initial perturbation this thread will
undergo a Plateau-Rayleigh instability and split into separate droplets \cite{Rayleigh1878}. In this case the computational domain is a rectangle spanning
$[-5,5]\times[-10,10]\times[-5,5]$ while the number of grid points is $101\times 201\times 101$ and the time step is $\Delta t=0.05$. Periodic boundary conditions are given in all directions.

\begin{figure}[!ht]
	\begin{center}
		\subfigure[t=0]{
			\includegraphics[width=1.5in]{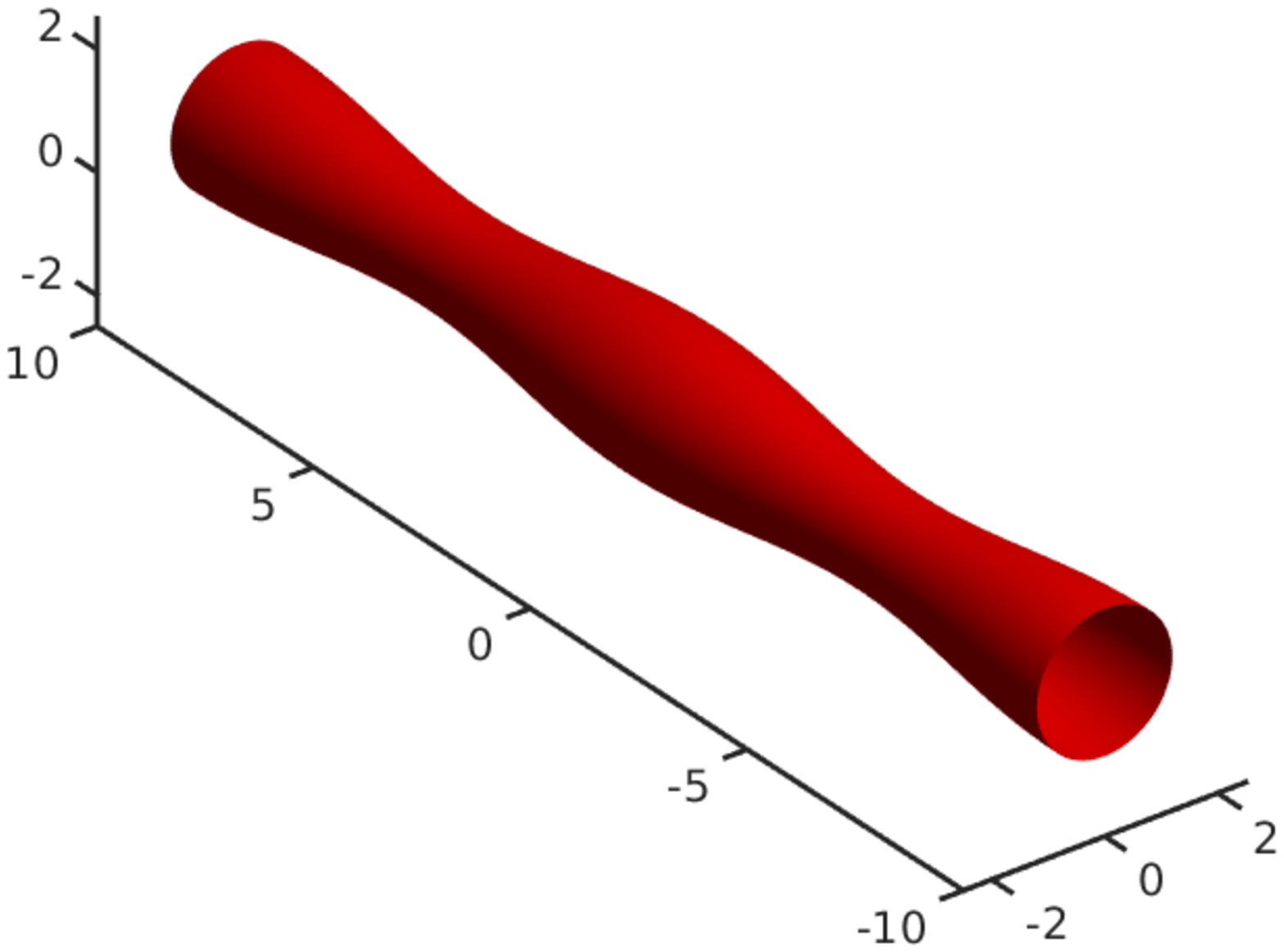}		
		}
		\subfigure[t=20]{
			\includegraphics[width=1.5in]{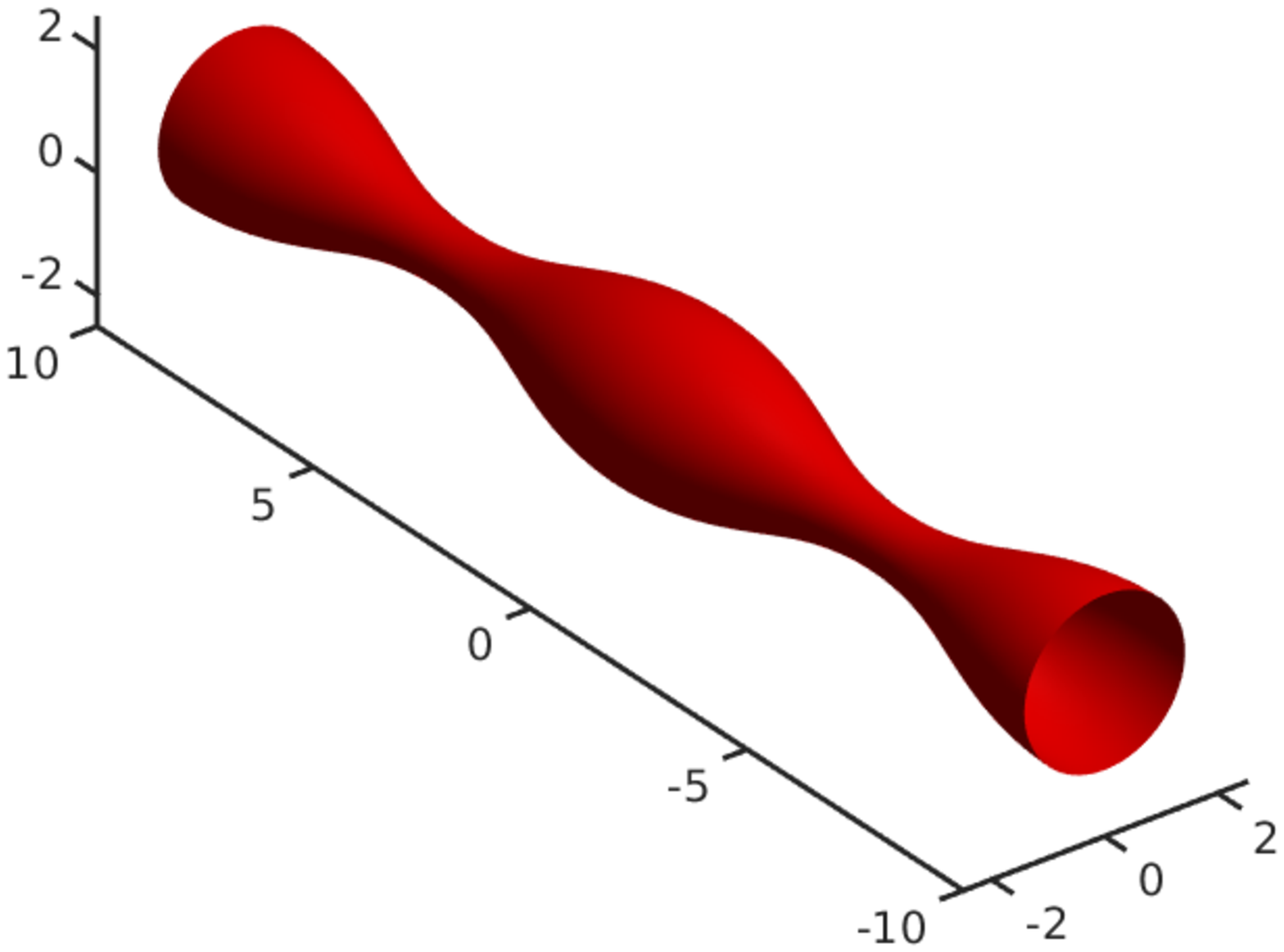}
		}\\
		\subfigure[t=28]{
			\includegraphics[width=1.5in]{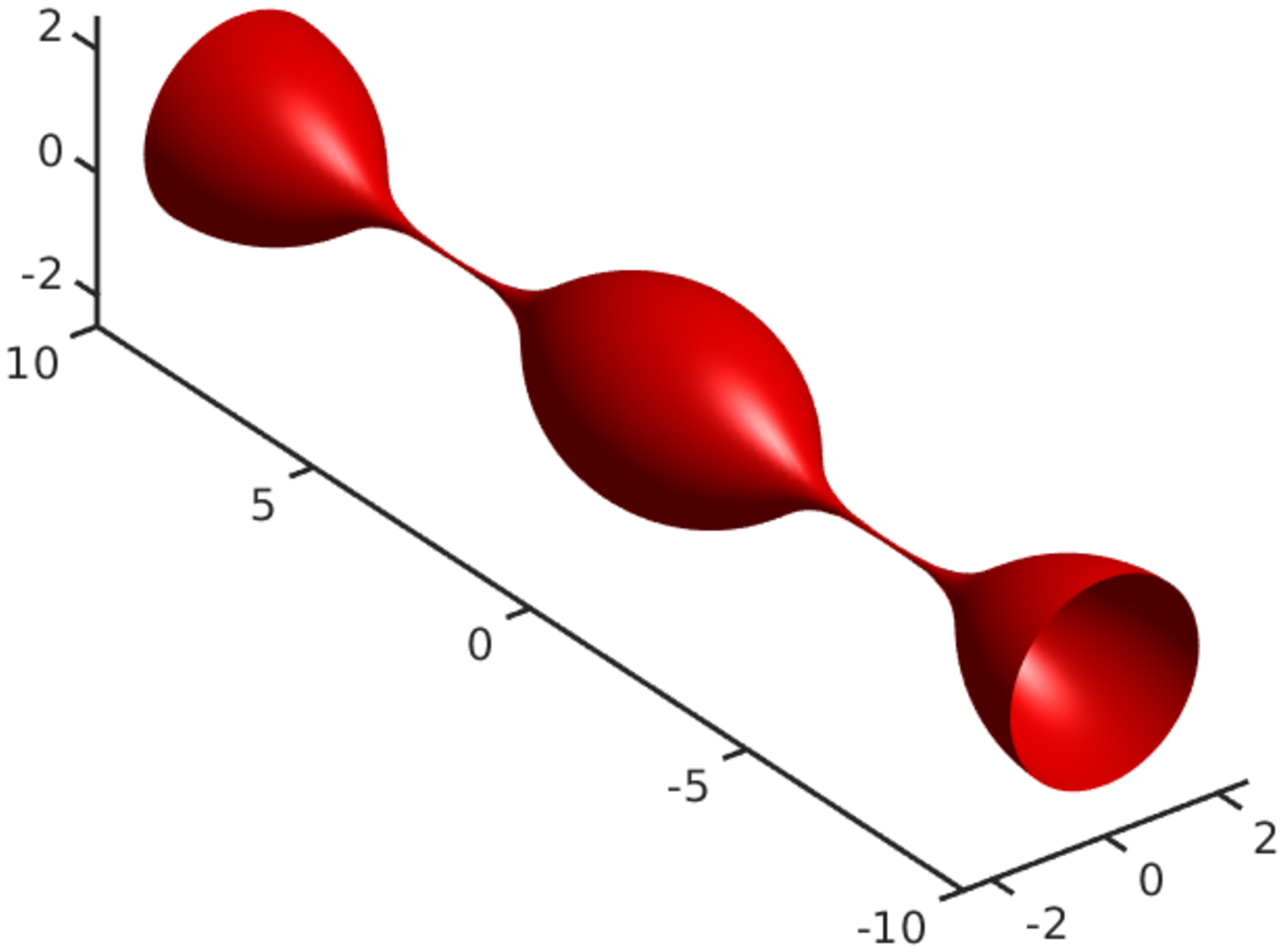}
		}
		\subfigure[t=28.5]{
			\includegraphics[width=1.5in]{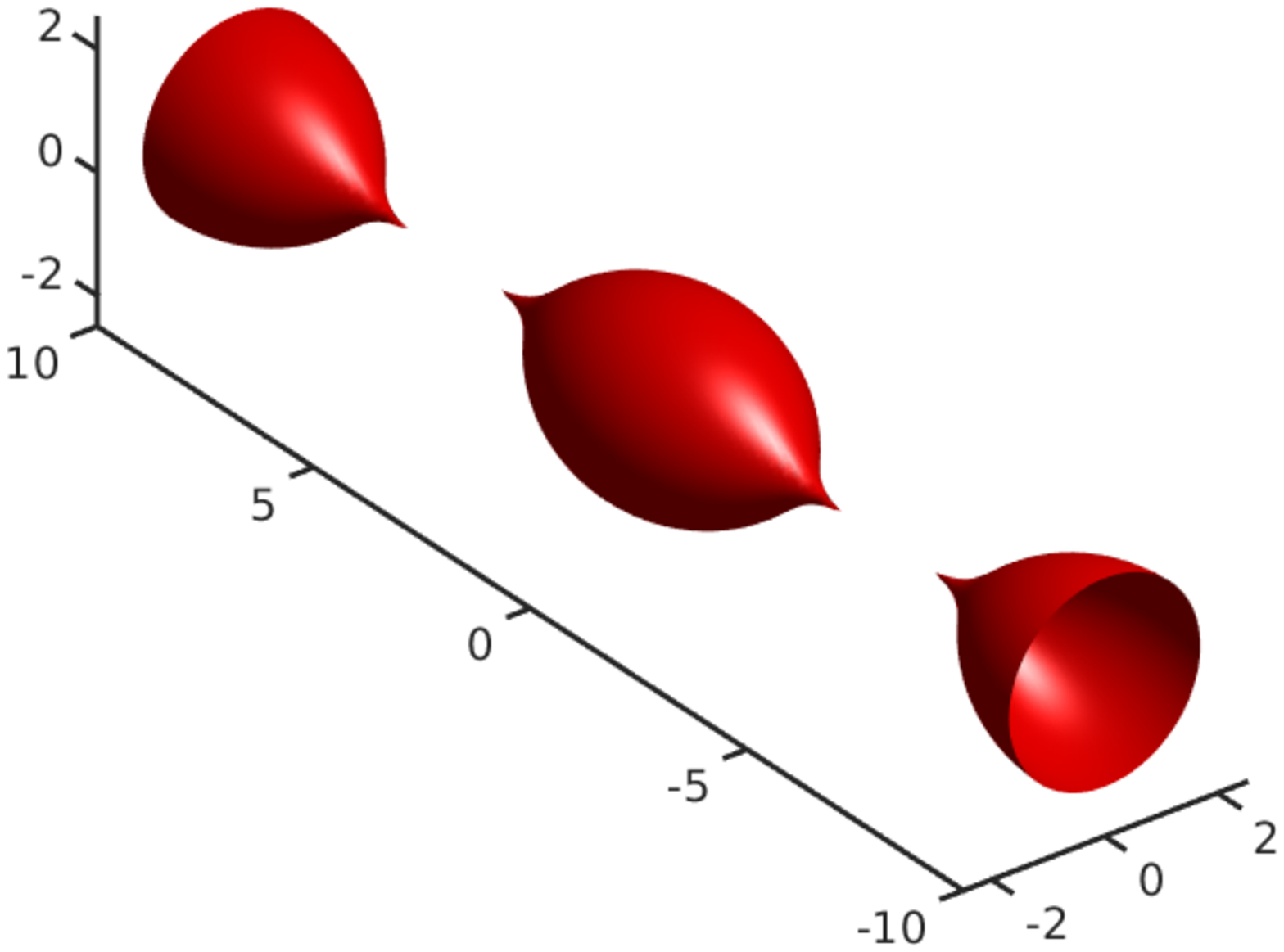}
		}\\
		\subfigure[t=30]{
			\includegraphics[width=1.5in]{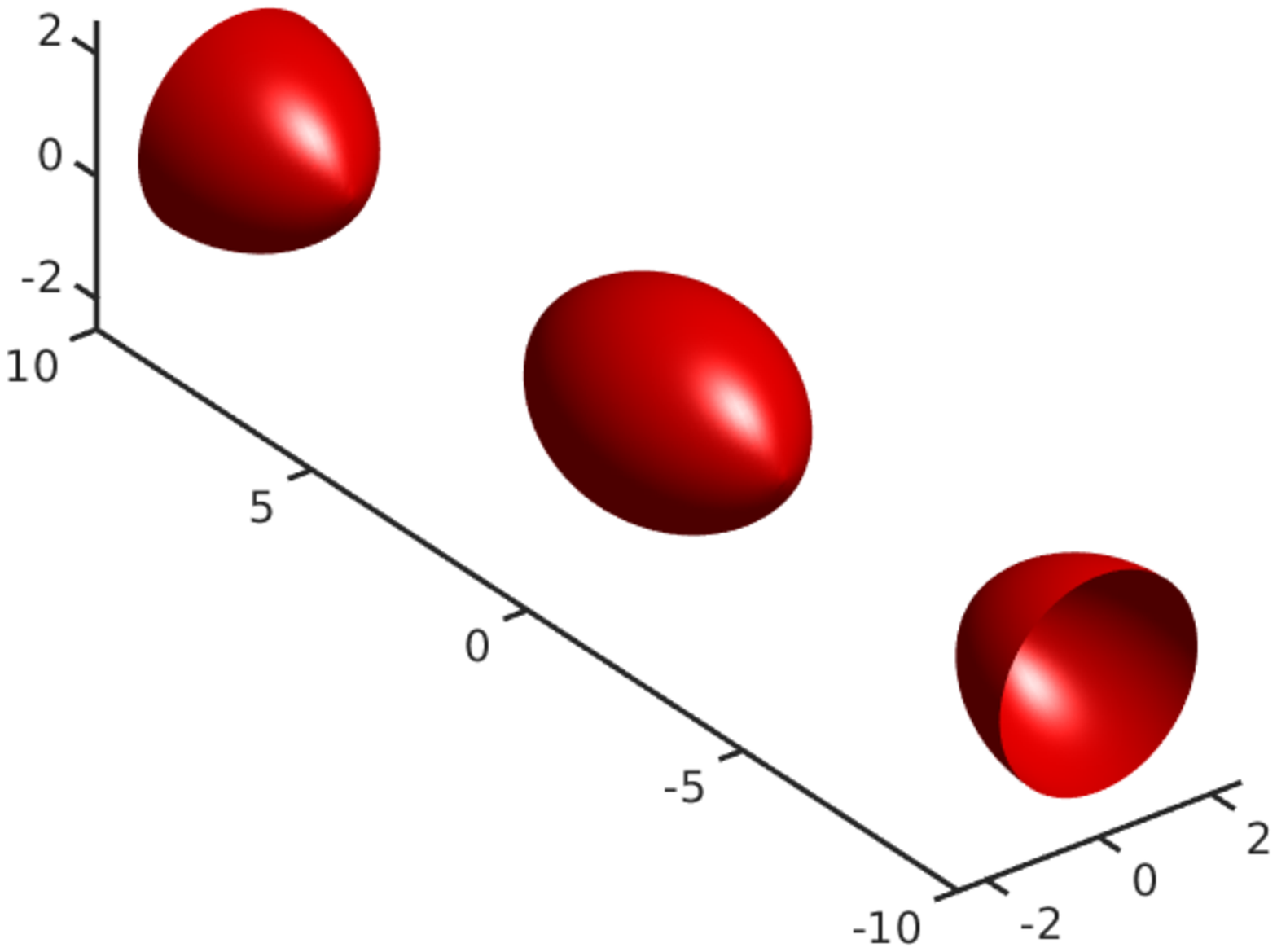}
		}
		\subfigure[t=100]{
			\includegraphics[width=1.5in]{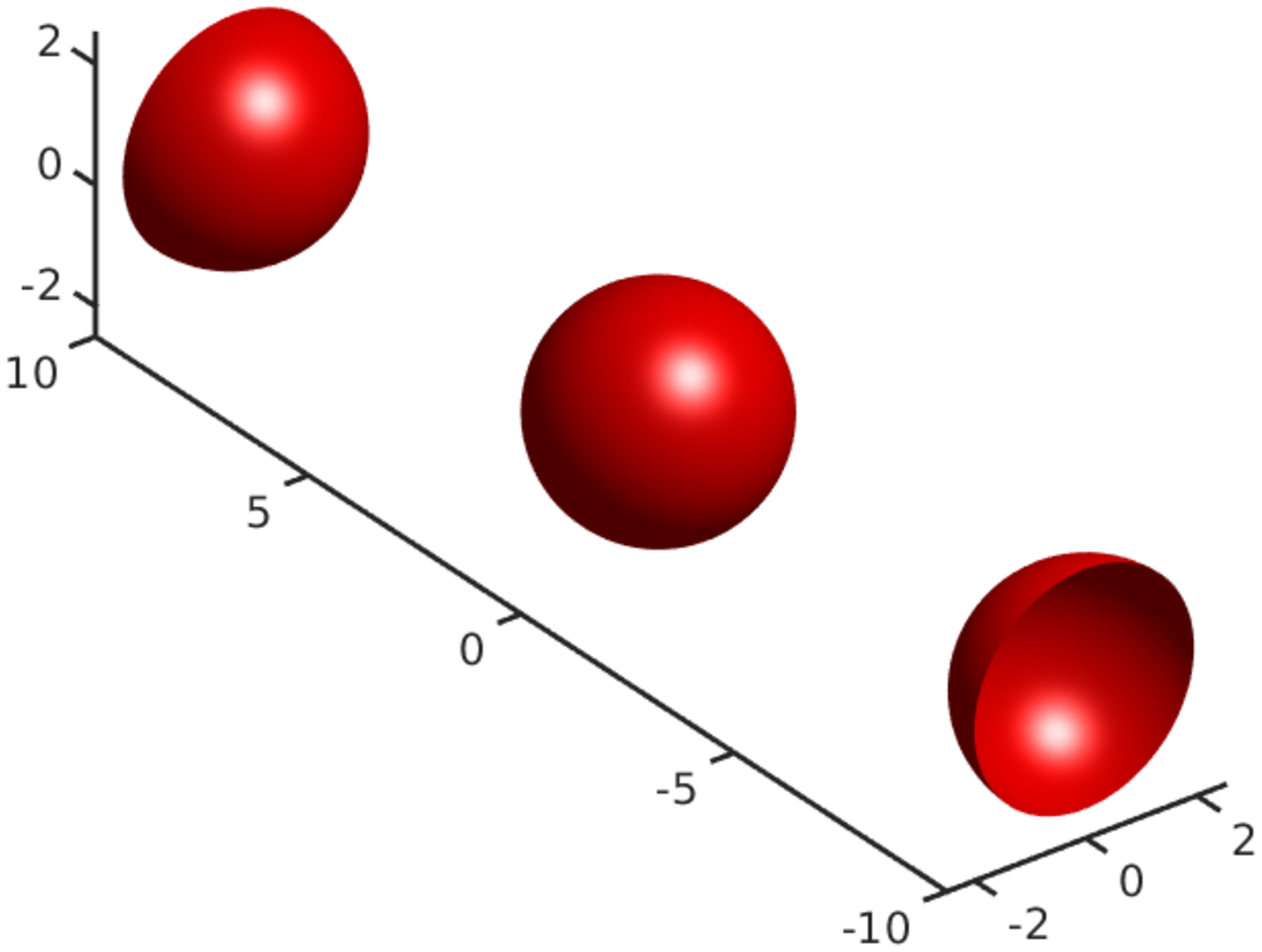}
		}
		\caption{Evolution of a fluid thread with a density ratio of $\lambda=10$, viscosity ratio of $\eta=10$, $\textrm{Re}=10$, $\textrm{We}=1$, and $\textrm{Fr}=\infty$. 
		Periodicity is assumed in all directions. With the given initial shape the droplet undergoes a Plateau-Rayleigh instability and breaks into spherical individual droplets.}
		\label{fig:PREvolution}
	\end{center}
\end{figure}

Initially a thread is aligned along the $y$-direction with an initial radius given by $1+0.2\cos(0.2\pi y)$ in the $x-z$ plane. 
It is then allowed to evolve under the conditions of 
$\textrm{Re}=10$, $\textrm{We}=1$, and $\textrm{Fr}=\infty$ with a density ratio of $\lambda=10$ and viscosity ratio of $\eta=10$.
The three-dimensional evolution of the interface is given in Fig. \ref{fig:PREvolution}. As time progresses the initially thin regions of the thread
become thinner. At a time of $t=28$ the thread is about to pinch off and extremely thin regions are observed between individual droplets, which are visible
at $t=28.5$. Over time the droplets become spherical to minimize the surface tension.

\begin{figure}[!ht]
	\begin{center}
		\subfigure[Volume]{
			\includegraphics[width=2.75in]{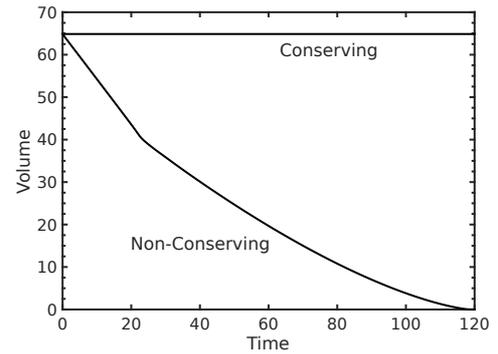}		
		}
		\subfigure[Volume Error]{
			\includegraphics[width=2.75in]{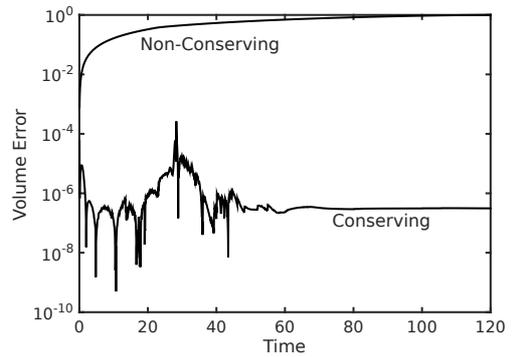}
		}		
		\caption{Comparison between volume preservation in the conserving and non-conserving schemes for a fluid thread undergoing a Plateau-Rayleigh instability
			with the initial condition shown in Fig. \ref{fig:PREvolution}.
			Due to volume loss the droplets in the non-conserving scheme disappear.}
		\label{fig:PRVolume}
	\end{center}	
\end{figure}

A comparison between the conserving and non-conserving schemes for the Plateau-Rayleigh instability is presented in Fig. \ref{fig:PRVolume}. As with the static bubble 
case the use of the non-conserving scheme results in complete volume loss. The conserving scheme has a maximum mass loss during the breakup observed
at approximately $t=28.5$, but quickly recovers to an error of less than $10^{-6}$.

\subsection{Droplet in Shear Flow}

As a final test consider the behavior of a droplet in the presence of shear flow. Under certain situations the droplet will extend and ``daughter" droplets may
form at the tip of the original droplet \cite{Li2000}. Here the computational domain spans $[-16,16]\times[-2,2]\times[-2,2]$ with a $513\times 65\times 65$ grid.
The time step is $\Delta t=0.025$. Periodic boundary conditions are taken in the $x-$ and $z-$ directions while wall-boundary conditions are taken in the $y-$direction.
The shear flow is imposed by specifying a velocity of $(y, 0, 0)$ on the $y=2$ and $y=-2$ walls.

\begin{figure}[!ht]
	\begin{center}
		\subfigure[t=0]{
			\includegraphics[width=2.75in]{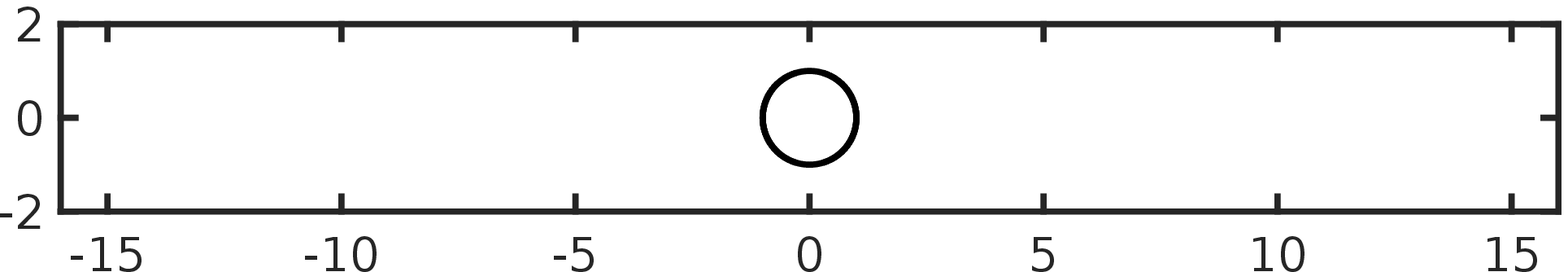}		
		}\\
		\subfigure[t=5]{
			\includegraphics[width=2.75in]{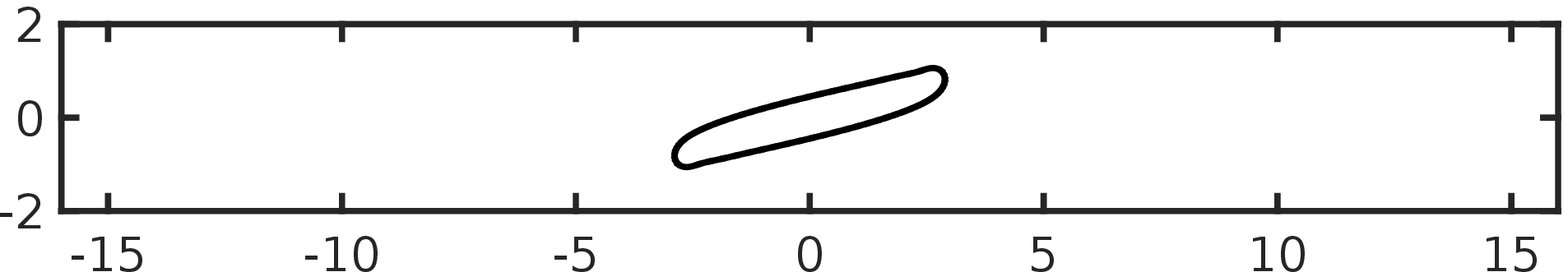}
		}\\
		\subfigure[t=10]{
			\includegraphics[width=2.75in]{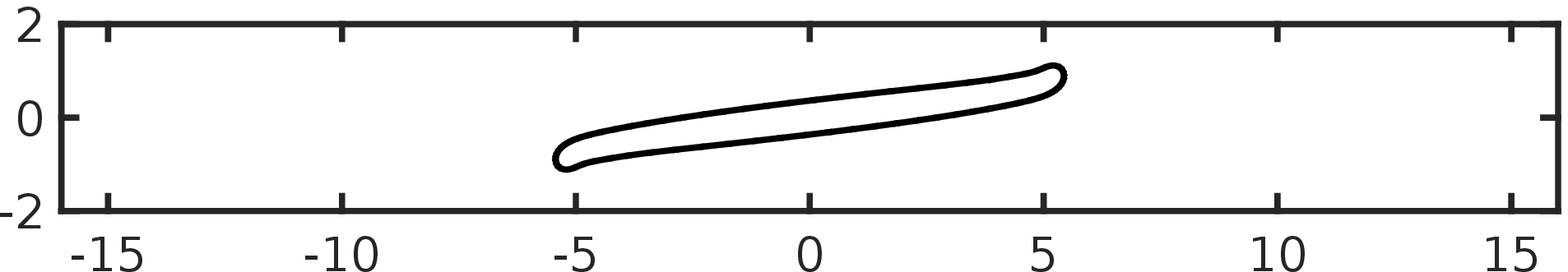}
		}\\
		\subfigure[t=15]{
			\includegraphics[width=2.75in]{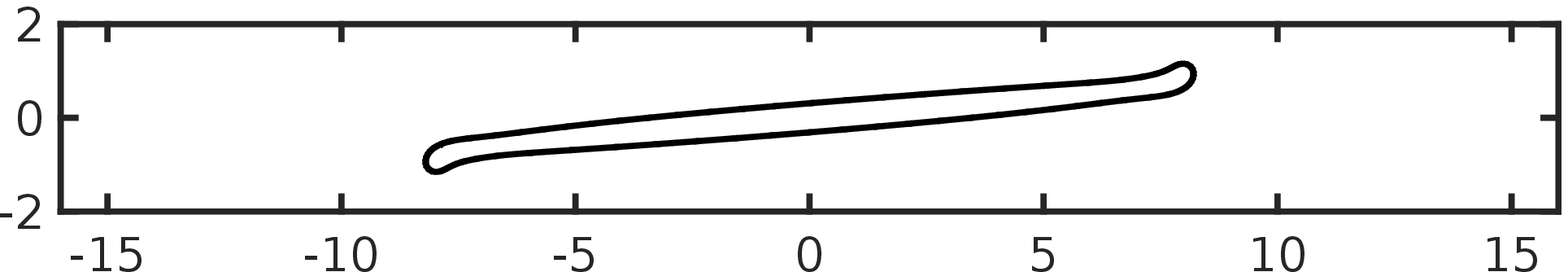}
		}\\	
		\subfigure[t=20.25]{
			\includegraphics[width=2.75in]{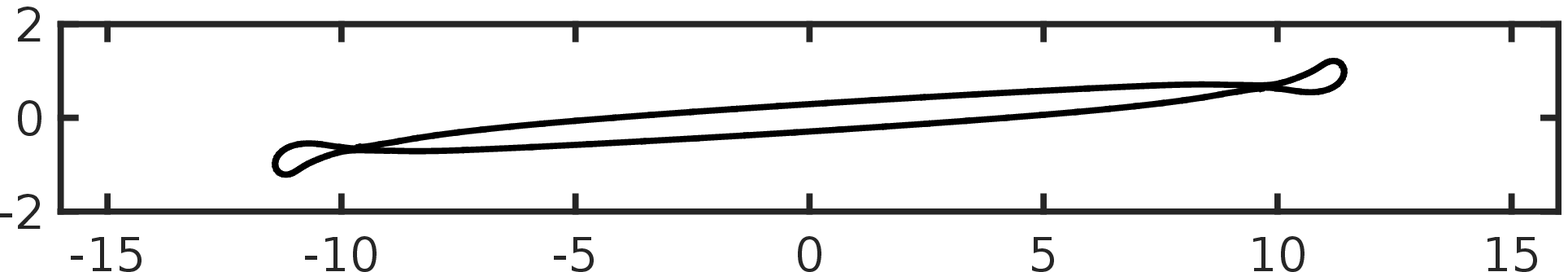}
		}	\\	
		\subfigure[t=20.5]{
			\includegraphics[width=2.75in]{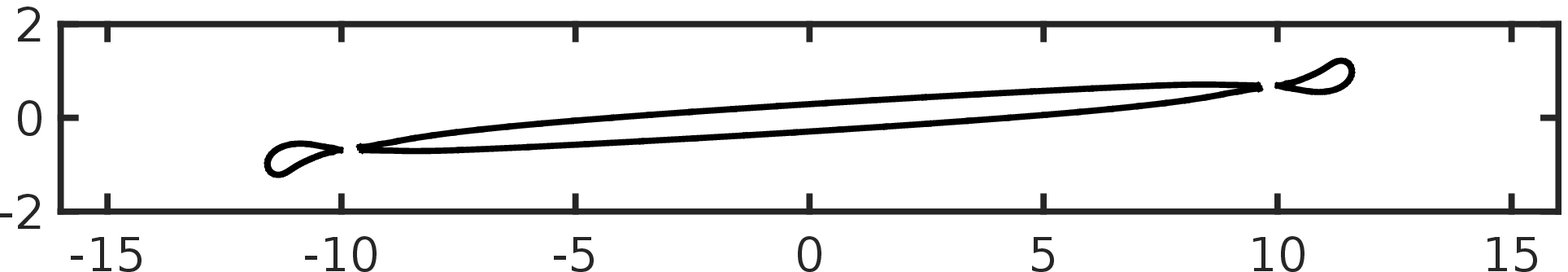}
		}
		\caption{Evolution of a droplet in the $x-y$ plane under shear flow. The density ratio is $\lambda=1$, the viscosity ratio is $\eta=1$, while $\textrm{Re}=1$,
		$\textrm{We}=1$, and $\textrm{Fr}=\infty$. The normalized shear rate is 1.}
		\label{fig:ShearEvolution}
	\end{center}
\end{figure}

The result for matched density and viscosity, $\lambda=1$ and $\eta=1$, with $\textrm{Re}=1$, $\textrm{We}=1$, and $\textrm{Fr}=\infty$ for an initially
spherical droplet of radius 1 is given in Fig. \ref{fig:ShearEvolution}. This figure shows the $x-y$ cross-section of the droplet up to the formation 
of the first ``daughter" droplets from the tips.

\begin{figure}[!ht]
	\begin{center}
		\subfigure[Volume]{
			\includegraphics[width=2.75in]{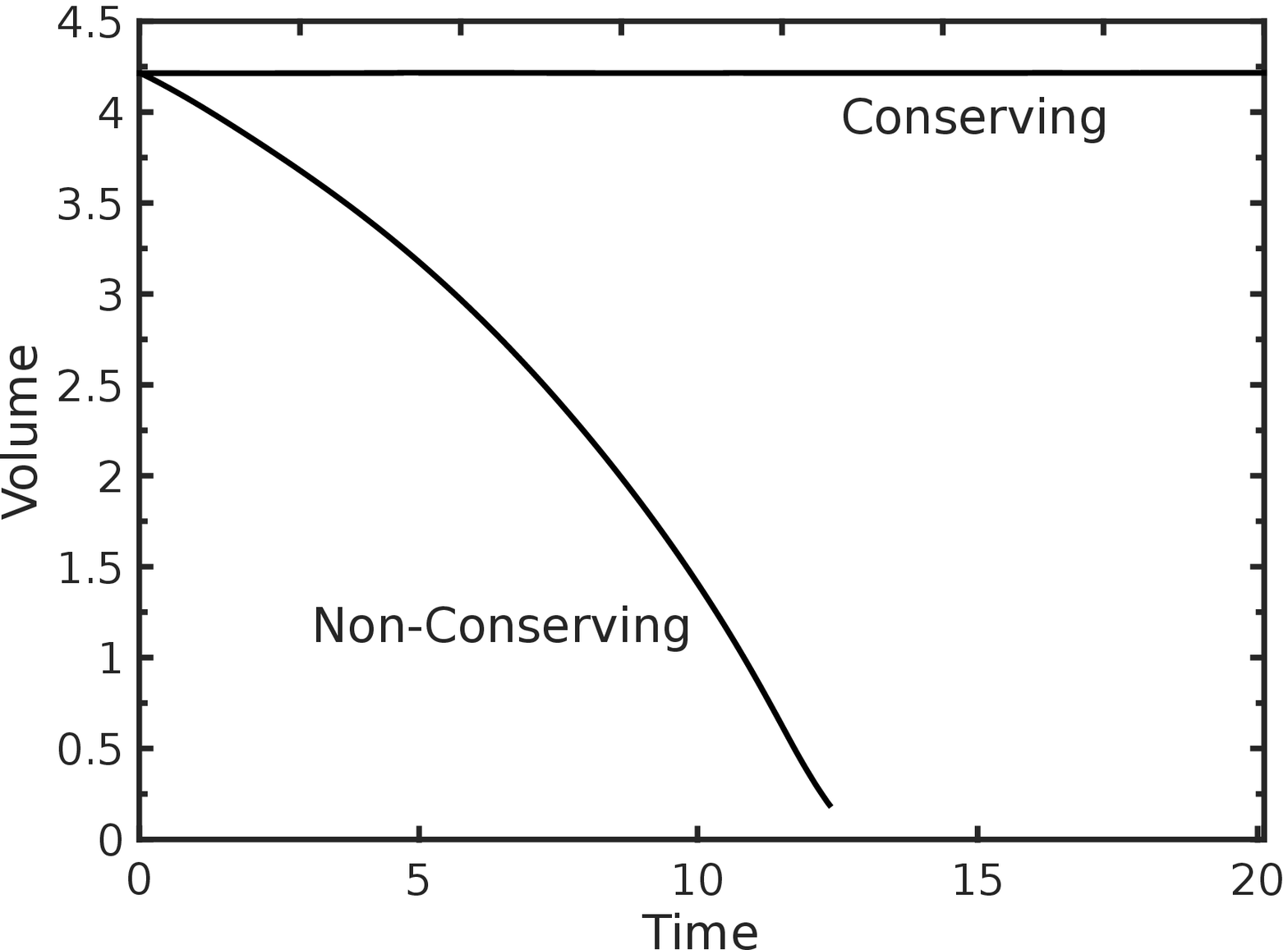}		
		}
		\subfigure[Volume Error]{
			\includegraphics[width=2.75in]{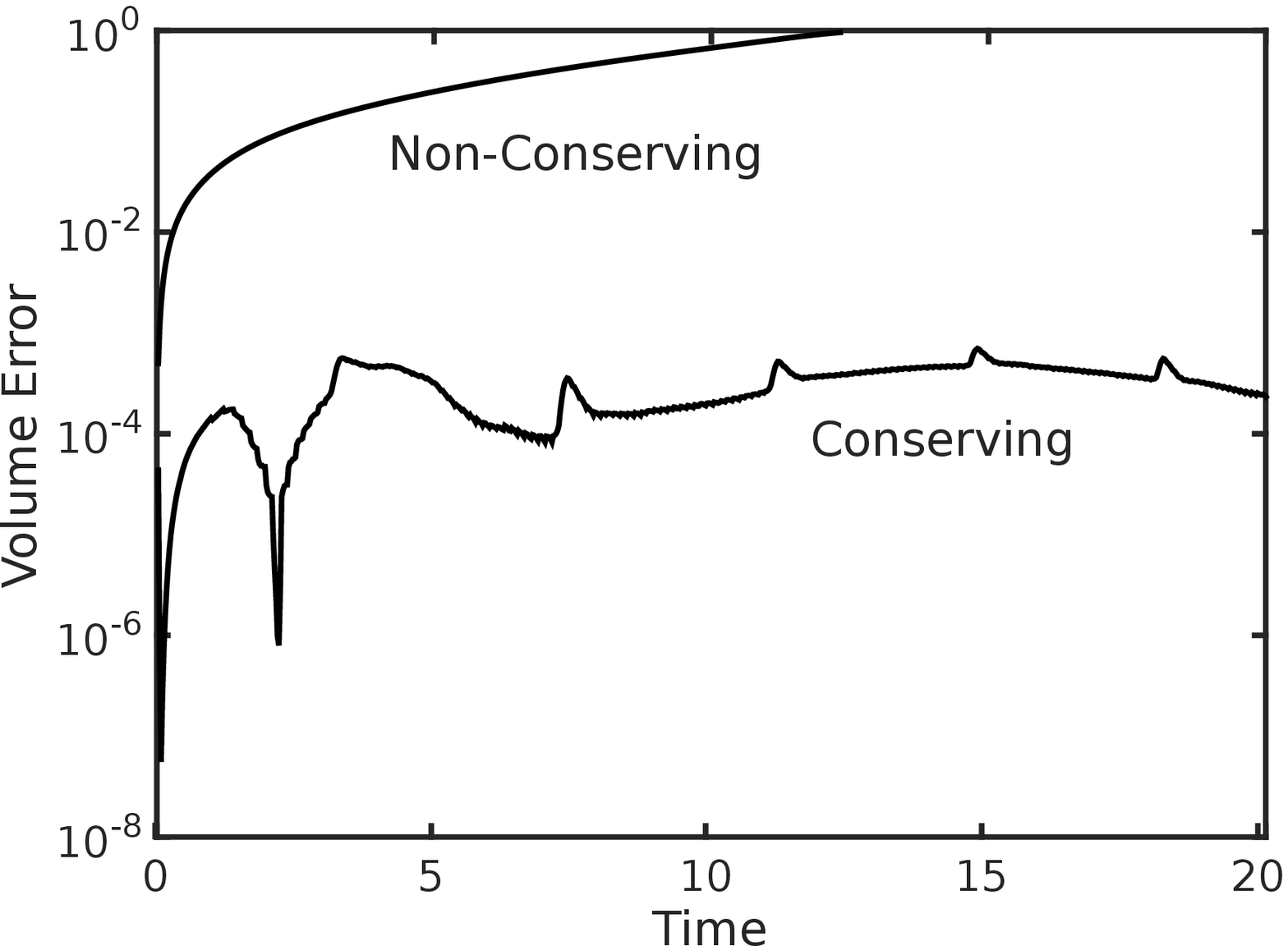}
		}		
		\caption{Comparison between volume preservation in the conserving and non-conserving schemes for the droplet in shear flow
			with the initial condition shown in Fig. \ref{fig:ShearEvolution}.
			Due to volume loss the droplet in the non-conserving scheme disappears.}
		\label{fig:ShearVolume}
	\end{center}
\end{figure}

Tracking the volume and volume error of the course of the simulation, Fig. \ref{fig:ShearVolume}, demonstrates results similar to the previous numerical tests. Over time the errors in the non-conserving
scheme accumulate and results in the complete disappearance of the droplet. The conserving scheme, on the other hand, is capable of ensuring that the mass errors do not grow
over time.

\section{Conclusion} 
\label{sec:conclusion}

In this work a new mass conserving projection method for multiphase Navier-Stokes systems has been presented. Various numerical tests have demonstrated that the 
proposed scheme is capable of conserving the mass in a wide variety situations, including a static bubble, a rising bubble, 
a fluid thread undergoing a Plateau-Rayleigh instability and a droplet in shear flow. 
Unlike many other mass correction schemes the material interface advects with the underlying fluid field. The general idea is applicable to any
system being investigated using a projection-based Navier-Stokes solver and is not limited to the particular discretization used here.

\section*{Acknowledgments}
	This work has been supported by the National Science Foundation through the Division of Chemical, Bioengineering, Environmental, and Transport Systems Grant \#1253739.

\end{document}